\documentclass[a4paper,12pt]{article}

\usepackage{amsmath,amssymb,amsthm,mathrsfs}
\usepackage{latexsym}
\usepackage[pdftex]{graphicx}
\numberwithin{equation}{section}

\newcommand{\be}{\begin{equation}}
\newcommand{\ee}{\end{equation}}
\newcommand{\bea}{\begin{eqnarray}}
\newcommand{\eea}{\end{eqnarray}}
\newcommand{\beann}{\begin{eqnarray*}}
\newcommand{\eeann}{\end{eqnarray*}}

\newcommand{\ba}{\begin{array}}
\newcommand{\ea}{\end{array}}

\newcommand{\del}{\partial}
\newcommand{\Dcal}{\mathcal{D}}

\newcommand{\lambdabar}{\bar{\lambda}}
\newcommand{\sigmabar}{\bar{\sigma}}
\newcommand{\xibar}{\bar{\xi}}
\newcommand{\Ocal}{\mathcal{O}}
\newcommand{\Zbar}{\bar{Z}}
\newcommand{\zbar}{\bar{z}}
\newcommand{\Yt}{\tilde{Y}}
\newcommand{\qbar}{\bar{q}}
\newcommand{\Fbar}{\bar{F}}
\newcommand{\psibar}{\bar{\psi}}
\newcommand{\Ybar}{\bar{Y}}
\newcommand{\chibar}{\bar{\chi}}
\newcommand{\mubar}{\bar{\mu}}

\newcommand{\Qbar}{\bar{Q}}

\newcommand{\alphadot}{\dot{\alpha}}
\newcommand{\Ncal}{\mathcal{N}}
\newcommand{\Zcal}{\mathcal{Z}}
\newcommand{\C}{\mathbb{C}}
\newcommand{\R}{\mathbb{R}}
\newcommand{\B}{{\mathcal B}}
\newcommand{\F}{{\mathcal F}}
\newcommand{\e}{\epsilon}
\DeclareMathOperator{\Tr}{Tr}

\DeclareMathOperator{\sign}{sign}

\newcommand{\D}{\mathscr{D}}
\newcommand{\A}{{\mathcal A}}

\newcommand{\et}{\tilde{\e}}
\newcommand{\Xibar}{\bar{\Xi}}
\newcommand{\sigmatilde}{\tilde{\sigma}}
\newcommand{\etat}{\tilde{\eta}}
\newcommand{\chit}{\tilde{\chi}}
\newcommand{\Vt}{\tilde{V}}
\newcommand{\Omegabar}{\bar{\Omega}}
\newcommand{\Yh}{\hat{Y}}
\newcommand{\sigmah}{\hat{\sigma}}

\begin{document}

\renewcommand{\thefootnote}{\fnsymbol{footnote}}
\begin{titlepage}

\vspace*{10mm}

\begin{center}
{\Large \textbf{
Coulomb Branch Localization\\[0.8em]
in\\[1.0em]
Quiver Quantum Mechanics
}}
\vspace*{12mm}

\normalsize{Kazutoshi Ohta\footnote{E-mail: kohta@law.meijigakuin.ac.jp} and Yuya Sasai\footnote{E-mail: sasai@law.meijigakuin.ac.jp}}

\vspace*{12mm}
\textit{
Institute of Physics, Meiji Gakuin University, Yokohama 244-8539, Japan  
}

\end{center}

\vspace*{15mm}

\begin{abstract}

We show how to exactly calculate the refined indices of $\Ncal=4$ $U(1)\times U(N)$ supersymmetric quiver quantum mechanics in the Coulomb branch by using the localization technique. 
The Coulomb branch localization is discussed from the viewpoint of
both non-linear and gauged linear sigma models.
A classification of fixed points in the Coulomb branch differs from one in the Higgs branch,
but the derived indices completely agree with the results which were obtained by the localization
in the Higgs branch. In the Coulomb branch localization, the refined indices can be written as a summation over different sets of the Coulomb branch fixed points.
We also discuss a space-time picture of the fixed points in the Coulomb branch.

\end{abstract}

\end{titlepage}

\newpage
\renewcommand{\thefootnote}{\arabic{footnote}}
\setcounter{footnote}{0}

\section{Introduction}

Some non-perturbative properties in supersymmetric theories are measured by indices \cite{Witten:1982df},
which are obtained by an evaluation of a partition function with periodic boundary conditions  in a compactified Euclidean time direction.
Recently, this kind of partition function is  exactly evaluated by using a localization method in various supersymmetric
theories. In \cite{Denef:2002ru}, it has been conjectured that
an index for multi-centered BPS black holes 
in a four-dimensional ${\cal N}{=}2$  supergravity theory
can be obtained from an  ${\cal N}{=}4$ supersymmetric quiver quantum mechanics.
The quiver quantum mechanics essentially appears as an effective theory for wrapped D-branes on different nontrivial cycles  in type II superstring theory which is compactified on a Calabi-Yau three manifold. Therefore,  the quiver quantum mechanics is expected to be a gauge theoretical description of the multi-centered BPS black holes \cite{Denef:2002ru,Denef:2007vg,Bena:2012hf,Lee:2012sc}.

The quiver quantum mechanics has two different effective descriptions,
 depending on two different  phases (branches) of the vacua \cite{Denef:2002ru}.
One of them  is called the Coulomb branch where some $U(1)$  gauge symmetries survive. In this branch, the D-particles (wrapped D-branes)  are located at different points in the bulk space-time.
Another is the Higgs branch where there is no gauge symmetry due to the Higgs mechanism. In this branch,  the D-particles are on top of each other.
These two branches are controlled by an external Fayet-Iliopoulos (FI) parameter $\zeta$ because the distance between those D-particles is proportional to $1/\zeta$. If $\zeta$ is small, the Coulomb branch description is reliable. But if $\zeta$ becomes large, the open strings stretching between the D-particles become tachyonic and the system goes to the Higgs branch.

In the Higgs branch, the theory reduces to a one-dimensional $\Ncal=4$ non-linear sigma model on the Higgs branch  moduli space ${\cal M}_H$, which is called quiver moduli space \cite{Denef:2002ru}.
The quantum mechanics on ${\cal M}_H$ gives rise to a Poincar\'e polynomial ($\chi_y$-genus) of ${\cal M}_H$,
which counts the number of the D-brane bound states in (a part of) the Calabi-Yau manifold, 
as a refined index.
The refined index of the quiver quantum mechanics in the Higgs branch can be derived
by using the localization technique.
In the Higgs branch localization, the path integral reduces  to residue integrals, which
can be  evaluated by the Jeffrey-Kirwan (JK) residue formula \cite{Hwang:2014uwa, Cordova:2014oxa, Hori:2014tda} or an equivalent moment map constraint analysis \cite{Ohta:2014ria}.
However, the Higgs branch localization also correctly reproduces the number of the BPS bound states in the supergravity
picture.
This is because the localization states that the partition function is locally independent of the coupling
constants including the FI parameter $\zeta$.

In this paper, we would like to revisit the exact evaluation of the refined index in the quiver quantum mechanics from
the viewpoint of the Coulomb branch picture.
The formula for the refined index of the multi-centered BPS black holes has been  derived by studying the moduli space of the supergravity solutions, which is known as the Manschot-Pioline-Sen (MPS) formula \cite{Manschot:2010qz, Manschot:2011xc}, but we here derive the refined
index from the original supersymmetric action of the quiver quantum mechanics
by using the localization technique.
As we explained above, the indices in  both of the Coulomb branch and Higgs branch apparently coincide with each other because of 
the independence of the coupling constants in the supersymmetric theory,
but we will show that the localization fixed points in the Coulomb branch and Higgs branch are totally different.
Despite the differences of the fixed points, the refined index coincides with each other as expected.
We  also see that  each fixed point in the Coulomb branch corresponds to the (localized) configuration of the BPS particles
in the space-time. This correspondence explains how the index (the number of the bound states)
appears in the supergravity (Coulomb branch) picture from the Higgs branch point of view.

To investigate the localization in the Coulomb branch, we concentrate on a simple quiver quantum mechanics, which has $U(1)\times U(N)$  gauge groups and  $k$ arrows.
This quiver quantum mechanics
is expected to describe BPS bound states of a magnetically charged BPS particle (monopole) with charge $k$ and $N$ electrically charged BPS particles (electrons) in the four-dimensional ${\cal N}{=}2$ supergravity theory \cite{Denef:2002ru}.
In the Higgs branch analysis, the number of bound states of this system is given by a binomial
coefficient ${k \choose N}$,
and the refined index becomes a $q$-binomial coefficient ${k \brack N}_{-y}$, where $y$ is a refined
parameter (fugacity) introduced by gauging  a global symmetry.
The $q$-binomial coefficient ${k \brack N}_{-y}$ can also be regarded as the Poincar\'e polynomial
of a Grassmannian $Gr(N,k)$, which is the Higgs branch moduli space of the $U(1)\times U(N)$
quiver quantum mechanics.
We will give an alternative derivation of this refined index in the Coulomb branch picture.

The organization of the paper is as follows:

In the succeeding section, we introduce a formulation of
one-dimensional supersymmetric non-linear sigma model with four supercharges,
and discuss some properties of the model. The essence of the non-linear sigma model
is contained in the single center case ($U(1)$ theory), but we also extend it to the multi-center case.
In section \ref{sec:loc}, we derive the localization formula for the supersymmetric non-linear sigma model.
With the periodic boundary condition for the Euclidean time, the partition function generally gives
the Atiyah-Singer index theorem on the Coulomb branch moduli. The index is evaluated by
an integration over zero modes associated with the flat directions of the Coulomb branch.
If we turn on the $\Omega$-background, the Coulomb branch moduli is lifted up and the path integral
is localized on isolated fixed points. Thus we can obtain the localization formula for the
refined index as a summation over the fixed points.
The refined index of the non-linear sigma model also agrees with a building block of the MPS formula
for Abelian nodes. We however need to consider further the indices from the original
linear sigma model (quiver quantum mechanics) in order to take 
non-Abelian
nodes into account.
In section \ref{sec:glsm}, we evaluate the refined index of the quiver quantum mechanics starting from the gauged linear sigma model.
With a careful treatment for the real auxiliary field,  we find an effective D-term condition after the chiral multiplets are integrated out. This effective D-term condition knows the fixed points for both of the Higgs branch and Coulomb branch. In fact,  if we take $\zeta/\beta \to \infty$ limit ($\beta$ is the periodicity of the Euclidean time), the moduli are localized at the Higgs branch fixed points, where the adjoint scalar moduli $\sigma_I$ $(I=1,\cdots, N)$ vanish. On the other hand, if we take  $\zeta/\beta \to 0$ limit,  $\sigma_I$ are  localized at $\sigma^\ast_I=\pm \frac{k}{2\zeta}$, which are the Coulomb branch fixed points. By taking the Coulomb branch limit,  the integrations over the gauge fields become trivial if the Coulomb branch fixed points are not degenerate.  If there are degenerate Coulomb branch fixed points, we encounter residue integrals over the gauge fields as in the case of the Higgs branch localization. In the Coulomb branch localization, the refined index can be written as a summation over different sets of the Coulomb branch fixed points. We compare our results with the MPS formula and discuss the correspondences. After taking the summation, we obtain the $q$-binomial coefficient. 
The final section is devoted to the summary and discussions.
In the appendix \ref{appconventions}, we review the $\Ncal=4$ $U(N)$ supersymmetric quantum mechanics according to \cite{Ohta:2014ria}. In the appendix \ref{sec:MPS}, we briefly summarize the MPS formula and write down some specific examples.

\section{1d ${\cal N}=4$ Supersymmetric Non-linear Sigma Model} \label{sec:nlsm}

In the IR limit of the quiver quantum mechanics, we expect that the effective theory in the Coulomb branch
is written in terms of the ${\cal N}=4$ supersymmetric non-linear sigma models of the vector multiplets.
If we have $n$ particles in the gravity side, we see that the gauge group of the model is $U(1)^n$.
So we first would like to discuss a formulation and general properties of the non-linear sigma model.
For simplicity, we will consider $U(1)$ gauge group case ($n=1$) for a while\footnote{
This case also describes the two-centered particle if we fix the center of motion and use a relative
coordinate between them.  
}.

The ${\cal N}=4$ vector multiplet in one dimension is obtained by a dimensional reduction from an ${\cal N}=1$ 
Abelian vector multiplet in
four dimensions, that is, the ${\cal N}=4$ vector multiplet contains a gauge field $A_\tau$, three scalar fields
$X^i$ ($i=1,2,3$), gauginos $(\lambda_\alpha,\bar{\lambda}_{\dot{\alpha}})$, and an auxiliary field $D$.
Following the notations and conventions in \cite{Wess:1992cp}
with the Euclidean signature, namely $\eta^{\tau\tau}=+1$,
the supersymmetric transformations of the fields are given by
\be
\begin{split}
\delta A_\tau &= -i \bar{\lambda}\bar{\sigma}_\tau \xi + i \bar{\xi}\bar{\sigma}_\tau \lambda,\\
\delta X_i &= -i \bar{\lambda}\bar{\sigma}_i \xi + i \bar{\xi}\bar{\sigma}_i \lambda,\\
\delta \lambda &= 2\sigma^{\tau i}\xi \dot{X}_i+i\xi D,\\
\delta D &= -\xi \sigma_\tau \dot{\bar{\lambda}}
-\dot{\lambda} \sigma_\tau \bar{\xi},
\end{split}
\label{SUSY trans}
\ee
where $\delta = \xi^\alpha Q_\alpha+\bar{\xi}_{\dot{\alpha}}\bar{Q}^{\dot{\alpha}}$ and $\dot{X}_i=\del_\tau X_i$.

In order to construct a generic action invariant under the above supersymmetric transformations,
it is useful to introduce the superspace formulation.
The superfield formulation of the generic supersymmetric quantum mechanics is discussed in \cite{Ivanov:1990jn,Diaconescu:1997ut}. In these formulations, the action is written in terms of the so-called linear multiplets
(see e.g. \cite{Gates:1983nr}), which is defined through the vector multiplet $V$ by
\be
\Sigma^i = -\frac{1}{2}\bar{D}_{\alphadot}\sigmabar^{i\alphadot \alpha}D_\alpha V,
\ee
where
\be
\begin{split}
D_\alpha & = \frac{\del}{\del \theta^\alpha}+i\sigma^\tau_{\alpha\dot{\alpha}}\bar{\theta}^{\dot{\alpha}}\del_\tau,\\
\bar{D}_{\dot{\alpha}} &= -\frac{\del}{\del \bar{\theta}^{\dot{\alpha}}}-i\theta^\alpha\sigma^\tau_{\alpha\dot{\alpha}}\del_\tau,
\end{split}
\ee
which obeys
\be
\begin{split}
& \{ D_\alpha, \bar{D}_{\dot{\alpha}} \} = - 2i \sigma^\tau_{\alpha\dot{\alpha}}\del_\tau, \\
& \{ D_\alpha, D_\beta \} = \{ \bar{D}_{\dot{\alpha}}, \bar{D}_{\dot{\beta}} \} = 0.
\end{split}
\ee
In terms of the components of fields, the linear multiplet is explicitly given by
\be
\begin{split}
\Sigma^i &= X^i +i \theta \sigma^i \bar{\lambda} + i \bar{\theta}\bar{\sigma}^i \lambda
+(\theta \sigma^i \bar{\theta})D\\
&\qquad
- \varepsilon^{ijk}(\theta \sigma_j \bar{\theta}) \dot{X}_k
 +(\theta\theta)(\bar{\theta}\bar{\sigma}^{\tau i}\dot{\bar{\lambda}})
+(\bar{\theta} \bar{\theta})(\theta\sigma^{\tau i}\dot{\lambda})
-\frac{1}{4}\theta\theta\bar{\theta}\bar{\theta}\ddot{X}^i.
\end{split}
\ee

The supersymmetric action, which contains quadratic order of velocities for the bosonic fields,
is given by an integration of a general function $K(\Sigma^i)$
of the linear multiplet over the superspace 
\be
\begin{split}
S_2 &= \int d\tau d^2\theta d^2\bar{\theta} \, K(\vec{\Sigma})\\
&=\frac{1}{2}\int d\tau \, \Big[
 G(\vec{X})(\dot{X}^i\dot{X}_i - D^2
+i\bar{\lambda}\bar{\sigma}^\tau\dot{\lambda}+i\lambda\sigma^\tau\dot{\bar{\lambda}})\\
&\qquad\qquad\quad
+\del_i G(\vec{X})(D \lambda\sigma^i\bar{\lambda}+\varepsilon^{ijk}\lambda \sigma_j\bar{\lambda}\dot{X}_k)
+\frac{1}{4}\del^2 G(\vec{X}) \lambda\lambda \bar{\lambda}\bar{\lambda}
\Big],
\end{split}
\label{SK}
\ee
where we have defined $G(\vec{X})\equiv \frac{1}{2}\del^2 K(\vec{X})$.
Note that the metric of the target space is always conformally
flat, that is,
\be
g_{ij} = G(\vec{X}) \delta_{ij}.
\ee
If this metric of the non-linear sigma model possesses the $SO(3)\simeq SU(2)$ isometry associated with the R-symmetry of ${\cal N}=4$ theory, $G$ needs to be a function of $|\vec{X}|$ only.

In addition to the kinetic part of the action (\ref{SK}), we can also introduce a first order action \cite{Denef:2002ru} by
\be
S_1 = \int d\tau \,
\Big[
U(\vec{X})D - \A_i(\vec{X})\dot{X}^i + \B_i(\vec{X})\bar{\lambda}\bar{\sigma}^i\lambda
\Big].
\ee
$S_1$ is invariant under the supersymmetric transformation (\ref{SUSY trans}) if a condition of
\be
\B_i = \del_i U = \varepsilon_{ijk}\del^j \A^k,
\ee
is satisfied.
A solution to this condition describes a monopole at the origin. 
This first order action is a one-dimensional analog of the supersymmetric Chern-Simons term in three dimensions,
and express the interaction with the bulk gauge field.

Now let us generalize the above to the multi-centered case.
We assume that the K\"ahler potential is a superposition of
functions of each relative coordinate, that is,
\be
K(\{\vec{\Sigma}\}) = \sum_{I\neq J}K_{IJ}(\vec{\Sigma}_{IJ}),
\ee
where we denote the linear multiplet of each particle
by $\vec{\Sigma}_I$ $(I=1,\ldots,n)$ and $\vec{\Sigma}_{IJ}\equiv \vec{\Sigma}_I - \vec{\Sigma}_J$.

Under this assumption, the kinetic part of the action is given by
\be
\begin{split}
S_2 &= \int d\tau d^2\theta d^2\bar{\theta} \, K(\{\vec{\Sigma}\})\\
&=\frac{1}{2} \sum_{I<J}\int d\tau \, \Big[
 G_{IJ}(\dot{\vec{X}}_{IJ} \cdot \dot{\vec{X}}_{IJ} - D_{IJ} D_{IJ}
+i\bar{\lambda}_{IJ}\bar{\sigma}^\tau \dot{\lambda}_{IJ}+i\lambda_{IJ}\sigma^\tau \dot{\bar{\lambda}}_{IJ})\\
&\qquad\qquad\quad
+\frac{\del G_{IJ}}{\del X^i_{IJ}}(D_{IJ} \lambda_{IJ}\sigma^i\bar{\lambda}_{IJ}
+\varepsilon^{ijk}\lambda^{IJ} \sigma_j\bar{\lambda}^{IJ}\dot{X}_{IJk})\\
&\qquad\qquad\quad
+\frac{1}{4}\frac{\del^2 G_{IJ}}{\del X_{IJ}^i \del X_{IJi}} 
\lambda_{IJ}\lambda_{IJ} \bar{\lambda}_{IJ} \bar{\lambda}_{IJ}
\Big],
\end{split}
\label{multi SK}
\ee
where $X^i_{IJ}=X^i_I-X^i_J$, etc.~and
\be
G_{IJ}(\vec{X}_{IJ}) = \frac{\del^2 K_{IJ}(\vec{X}_{IJ})}{\del X^i_{IJ} \del X_{IJi}}.
\ee

We can also introduce the interaction term with the bulk gauge fields
\be
S_1 = \int d\tau
\left[
U_I(\{\vec{X}\}) D^I
-{\cal A}^I_i (\{\vec{X}\})\dot{X}^i_I
+{\cal B}_{iIJ}(\{\vec{X}\})\bar{\lambda}^I\bar{\sigma}^i \lambda^J
\right].
\ee
To preserve the supersymmetry, each field should satisfy
\be
{\cal B}_{iIJ}=\frac{\del U_J}{\del X^{iI}}=\varepsilon_{ijk}\frac{\del {\cal A}^{kJ}}{\del X^{jI}},
\ee
and $\del_{iI}U_J=\del_{iJ}U_I$. If we assume that  each potential $U_I$ is a superposition of
the functions of the relative distance of the particles, we have a solution to the above condition by
\be
U_I(\{\vec{X}\}) =  \sum_J \frac{k_{IJ}}{2|\vec{X}_{IJ}|}-\zeta_I,
\ee
where $k_{IJ}$ and $\zeta_I$ are integral constants corresponding to
the number of the arrows between nodes $I$ and $J$ (Dirac-Schwinger-Zwanziger (DSZ) product of the charges)
and the FI parameter of $I$-th node (phase of the central charge),
respectively.

\section{Localization}  \label{sec:loc}

\subsection{Index theorem}

Let us first discuss the two-centered particle case. If we remove the center-of-mass motion and denote
the relative coordinate by a single linear multiplet, this is the case of $n=1$ in the previous formulation.

Now if pick up a single supercharge from four supercharges in the system, by 
a linear combination $Q=\frac{i}{\sqrt{2}}(Q^1 -\bar{Q}^{1})$,
the transformation law of the vector multiplet
with respect to this $Q$ becomes
\be
\begin{array}{lcl}
Q A_\tau = -\eta^3, &&\\
Q X^i = i \eta^i, && Q\eta^i = i \dot{X}^i,\\
Q D = i \dot{\chi}, && Q \chi = i D,
\end{array}
\ee
where we have defined
\be
\begin{array}{lll}
\eta^1= \frac{i}{\sqrt{2}}(\lambda_2+\bar{\lambda}_{2}),
& \eta^2 = \frac{1}{\sqrt{2}}(\lambda_2-\bar{\lambda}_{2}),
& \eta^3 = \frac{i}{\sqrt{2}}(\lambda_1+\lambda_{1}).
\end{array}
\ee
Note that $Q^2=-\del_\tau$, which is an isometry translation along $\tau$.
Using the supercharge $Q$, we can write the action in the $Q$-exact form
in the $A_\tau=0$ gauge
\be
\begin{split}
S_2 &= -\frac{i}{2}Q \int d\tau \,
\Big[
G(\vec{X})(\eta^i \dot{X}_i-\chi D)
-\frac{1}{2}\chi \varepsilon^{ijk}\del_i G \eta_j \eta_k
\Big]
\\
&=\frac{1}{2}\int d\tau \, \Big[
G(\dot{X}^i \dot{X}_i
- \eta_i \nabla_\tau^{ij} \eta_j
-\chi \dot{\chi}
-D^2)\\
&\qquad\qquad\quad
-\frac{1}{2}D\varepsilon^{ijk}\del_i G \eta_j \eta_k
+\chi \varepsilon^{ijk}\del_i G \dot{X}_j \eta_k
+\del^2 G \chi \eta^1 \eta^2 \eta^3
\Big],
\end{split}
\ee
and
\be
\begin{split}
S_1 &= -i Q \int d\tau \,
\Big[
U(\vec{X})\chi-i\A_i(\vec{X})\eta^i
\Big]\\
&=\int d\tau \, \Big[
U D - i\A_i \dot{X}^i
- \del_i U \chi \eta^i
-\frac{i}{2}\F_{ij}\eta^i \eta^j
\Big],
\end{split}
\ee
where $\nabla_\tau^{ij}=\delta^{ij}\del_\tau-G^{-1}\del^i G \dot{X}^j$
and $\F_{ij}=\del_i \A_j - \del_j \A_i = \varepsilon_{ijk}\del^k U$.
Thus we find that the partition function is independent of the couplings and exact at the 1-loop level
because of the $Q$-exactness of the action.

We so far have not specified the coordinates of the three dimensional space. Let us now take the
 spherical coordinates $\vec{X}=(r,\theta, \phi)$. Then we also have fermions
$(\eta^r,\eta^\theta,\eta^\phi)$ associated with the spherical coordinates.
Assuming that the conformally flat
metric $G$ and the potential $U$ are spherically symmetric, that is, $G$ and $U$ are  functions of $r$ only,
and the derivatives of $G$ and $U$ do not vanish in general, then we find that the bosonic fields $r$ and $D$,
and corresponding fermions $\eta^r$ and $\chi$ are massive
due to the potential $U(r)$,
while $(\theta,\phi)$ and $(\eta^\theta,\eta^\phi)$ are massless.
This means that there are flat directions along $(\theta,\phi)$-coordinates and the supersymmetric multiplet
contains the zero (constant) modes. We also note that the non-zero components of the external field strength
is $\F_{\theta\phi}$ only under this assumption.

Introducing now two couplings for  the action $S_1$ and $S_2$, let us consider the total action
\be
S = tS_2 + t' S_1.
\ee
We can expect that the partition function is independent of the couplings $t$ and $t'$.
Utilizing this coupling independence, we can consider the limit of $t'\gg 1$ without changing the value of
the partition function. In this limit, we obtain the path integral
\be
Z \simeq \int \D^3 \vec{\Sigma} \, e^{-t'S_1}.
\ee
After integrating over all fields except for $r$ and possible zero modes, we obtain
\be
\begin{split}
Z &\simeq  \int \D r d\theta_0 d\phi_0 d\eta_0^\theta d\eta_0^\phi \, \delta(U(r))U'(r)
e^{-\frac{it' \beta}{2}F_{\theta \phi}\eta_0^\theta \eta_0^\phi}\\
&=\sum_i \int \D r d\theta_0 d\phi_0 d\eta_0^\theta d\eta_0^\phi \, \delta(r-r_i^*)\sign (U'(r_i^*))
e^{-\frac{it' \beta}{2}F_{\theta \phi}\eta_0^\theta \eta_0^\phi},
\end{split}
\ee
where $U'(r)=\del_r U(r)$, $r_i^*$'s are solutions (zeros) to the equation $U(r)=0$,
and $\theta_0$, $\phi_0$, $\eta_0^\theta$ and $\eta_0^\phi$ represent the zero modes.

This result says that we can replace $e^{-S_1}$ by
\be
e^{-S_1} = \sum_i \delta(r-r^*_i)\sign(U'(r^*_i))
\delta(\tilde{\theta})\delta(\tilde{\phi}) \delta(D)
\delta(\tilde{\eta}^\theta)\delta(\tilde{\eta}^\phi)\delta(\chi)
e^{-\frac{i}{2}\int d\tau \, F_{\theta\phi}\eta^\theta \eta^\phi}.
\ee
Plugging this back to the original path integral, we finally find
\be
Z = -\frac{i\beta}{2}\sum_i \sign(U'(r^*_i)) \int_{r=r_i^*} d\theta_0 d\phi_0 \, F_{\theta\phi}.
\ee
In particular, if we consider a bound state of one electron and one monopole with a monopole charge $k$,
the Coulomb potential is given by
\be
U(r) = \frac{k}{2r} -\zeta.
\ee
Then we have a solution $r^* = k/2\zeta$
 if $k$ and $\zeta$ have the same sign. Using $\frac{1}{4\pi}\int d\theta d\phi F_{\theta\phi}=k$
 and $\sign U'(r^*)=-\sign k$, we get
$Z=2\pi i \beta |k|$ as the index\footnote{We can include the factor $2\pi i$ into the normalization of the path integral measure.}.

In the above arguments, we find that the path integral is localized at a surface of the fixed radius $r=r^*_i$,
which is a solution to $U(r)=0$.
For multi-centered particle, this surface is the moduli space of the Coulomb branch. So repeating the above
argument, we generally obtain the Atiyah-Singer index on
the Coulomb moduli space ${\cal M}_C$ \cite{Manschot:2011xc}
\be
Z = \int_{{\cal M}_C} \left. \hat{A}(T{\cal M}_C){\rm ch}(F)\right|_{\rm vol},
\ee
where 
$\hat{A}(T{\cal M}_C)$ is the Dirac genus and
${\rm ch}(F)$ is the total Chern character of the two-form field strength $F$, which is induced by the interaction with the bulk gauge field.


\subsection{Turning on the $\Omega$-background}

As we have seen, the path integral of the partition function finally reduces to integrations
over the moduli space of the Coulomb branch, which is parametrized by the massless zero modes.
After integrating over the moduli, we obtain the Atiyah-Singer index on the Coulomb branch,
but an explicit evaluation of the index is difficult in general.

We however can utilize the localization technique more for this system,
since we are constructing the BRST exact action for the non-linear sigma model on the Coulomb branch.
The zero modes in the residual integral correspond to fixed surfaces of the BRST transformation.
So, in order to make the fixed surface to be isolated fixed points, we can turn on masses for
the massless modes.
This can be done by ``gauging'' a global R-symmetry. It is also equivalent to
an introduction of the so-called $\Omega$-background \cite{Nekrasov:2003rj}.

Originally, the ${\cal N}=4$ supersymmetric non-linear sigma model has  $SU(2)_J$ R-symmetry,
which is an isometry on the Coulomb branch. If we define  complex fields by
\be
\begin{array}{ll}
Z = X^1-iX^2, & \bar{Z}=X^1+iX^2,\\
\lambda_z = \eta^1-i\eta^2, & \lambda_{\bar{z}} = \eta^1+i\eta^2,
\end{array}
\ee
then a $U(1)$ part of the $SU(2)_J$ acts on these fields by
\be
Z\to e^{i\theta_J} Z, \quad \lambda_z \to e^{i\theta_J}\lambda_z,
\ee
etc. The gauging of the R-symmetry means that the derivatives in the BRST transformation and the action
are modified to the covariant derivatives with a constant background of a $U(1)$ gauge field
\be
\del_\tau Z\to (\del_\tau + i\e) Z, \quad \del_\tau \lambda_z \to (\del_\tau + i\e)\lambda_z.
\ee
As a consequence, we have a modified BRST transformation
\be
\begin{array}{lcl}
Q_\e Z = i \lambda_z, && Q_\e \lambda_z = i(\del_\tau + i\e) Z,\\
Q_\e \bar{Z} = -i \lambda_{\bar{z}}, && Q_\e \lambda_{\bar{z}} = -i(\del_\tau - i\e) \bar{Z},\\
Q_\e A = i\eta, &&\\
Q_\e \sigma = \eta, && Q_\e \eta = - \del_\tau \sigma,\\
Q_\e D = i \dot{\chi}, && Q_\e \chi = i D,
\end{array}
\ee
where we have defined $A=A_\tau$, $\sigma=X^3$ and $\eta=i\eta^3$, to make them coincide with later conventions
(See also Appendix \ref{appconventions}). The kinetic part of the action
is still written by an exact form with respect to $Q_\e$
\be
\begin{split}
S_2^\e &= \frac{i}{2}Q_\e \int d \tau \bigg[
G(Z,\bar{Z},\sigma)\left(\frac{1}{2}\lambda_{\bar{z}} (\dot{Z}+i\e Z)
-\frac{1}{2}\lambda_z (\dot{\bar{Z}}-i\e \bar{Z})-i\eta \dot{\sigma} + \chi D\right)\\
&\qquad\qquad\qquad\qquad+i \chi \del_z G \lambda_{\bar{z}} \eta
+i\chi \del_{\bar{z}} G \eta \lambda_z
+\chi\del_\sigma G \lambda_z \lambda_{\bar{z}}
\bigg].
\end{split}
\ee
Similarly, for the potential and interaction part, we have
\be
S^\e_1 = -i Q_\e \int d\tau \,
\Big[
U(Z,\bar{Z},\sigma)\chi-i\A_z \lambda_{z}+i\A_{\bar{z}} \lambda_{\bar{z}}-\A_\sigma \eta
\Big].
\ee
Using the explicit monopole potential, the bulk gauge field is explicitly given by
\be
\A_z = -\frac{ik}{4}\left(\frac{\sigma}{r}\mp 1\right)\frac{\bar{Z}}{|Z|^2},\quad
\A_{\bar{z}} = \frac{ik}{4}\left(\frac{\sigma}{r} \mp 1\right)\frac{Z}{|Z|^2},\quad
\A_\sigma =0.
\label{monopole gauge}
\ee
The sign in front of 1 corresponds to a choice of the bulk gauge field. The gauge of the minus and plus sign
is regular at $Z=\bar{Z}=0$ if $\sigma>0$ and $\sigma<0$, respectively.

Due to the $\Omega$-background $\e$, $Z$ and $\bar{Z}$, which represent positions of the superparticles
on $(x^1,x^2)$-plane, become massive. The BRST fixed point equation also says $Z=\bar{Z}=0$ at
the fixed (saddle) point of the WKB approximation. The superparticle coordinate $\sigma$ is also massive
as well as the previous case and fixed by the moment map constraint $U(\sigma)=0$.
So there is no more integration over the massless modes (zero modes) in the path integral.

Expanding now all fields around a fixed point like
\be
Z = 0 + \frac{1}{\sqrt{t}}\tilde{Z},\qquad
\bar{Z} = 0+\frac{1}{\sqrt{t}}\tilde{\bar{Z}},\qquad
\sigma = \sigma^* + \frac{1}{\sqrt{t}}\tilde{\sigma},\qquad
D = 0 + \frac{1}{\sqrt{t}} \tilde{D},
\ee
for bosons and
\be
\lambda_z = 0 + \frac{1}{\sqrt{t}} \tilde{\lambda}_z,\qquad
\lambda_{\bar{z}} = 0 + \frac{1}{\sqrt{t}} \tilde{\lambda}_{\bar{z}},\qquad
\eta = 0 +\frac{1}{\sqrt{t}} \tilde{\eta},\qquad
\chi = 0 + \frac{1}{\sqrt{t}} \tilde{\chi},
\ee
for fermions, where {\it tilde} denotes fluctuations rescaled thanks to the invariance of the supersymmetric measure
and $\sigma^*$ is a constant solution to $U(\sigma)=0$,
the kinetic part of the action is expanded up to the quadratic order of the fluctuations
\be
\begin{split}
t S^\e_2 &=\frac{1}{2}\int d\tau \,
G(\sigma^*)\Big(
|\dot{\tilde{Z}}+i\e \tilde{Z}|^2+\dot{\tilde{\sigma}}^2 -\tilde{D}^2\\
&\qquad\qquad\qquad
+\frac{1}{2}\tilde{\lambda}_{\bar{z}}(\dot{\tilde{\lambda}}_z+i\e \tilde{\lambda}_z)
+\frac{1}{2}\tilde{\lambda}_z(\dot{\tilde{\lambda}}_{\bar{z}}-i\e\tilde{\lambda}_{\bar{z}})
-\tilde{\eta}\dot{\tilde{\eta}}+\tilde{\chi}\dot{\tilde{\chi}}
\Big)+{\cal O}(1/\sqrt{t}).
\end{split}
\ee

For the first order interaction part,
it is necessary to pay a little attention to the localization.
Firstly, the potential, which is coupled with the D-field in the first order action,
will gives a constraint $U=0$ after eliminating the auxiliary D-field.
So we need to incorporate terms associated with the potential
into the evaluation of the 1-loop determinant.
Expanding these terms up to the quadratic order, we find
\be
-it Q_\e \int d\tau \,  U\chi=\int d\tau \,
\Big[
U'\tilde{\sigma}\tilde{D}
+ U'\tilde{\chi}\tilde{\eta}
\Big]+{\cal O}(1/\sqrt{t}),
\ee
where $U' \equiv \left.\frac{\del U}{\del \sigma}\right|_{Z=\bar{Z}=0,\sigma=\sigma^* }$.
Secondly, since the interaction term with the bulk gauge field is now pure imaginary and it is a phase in the path integral, we evaluate them at the fixed points just as a $Q_\e$-closed operator,
like the supersymmetric Chern-Simons term in the Euclidean three-dimensional space-time
\cite{Ohta:2012ev}. Using the gauge (\ref{monopole gauge})
of the minus sign, we find at the fixed point
\be
\left.
-Q_\e \int d\tau \,
\left[
\A_z \lambda_z - \A_{\bar{z}}\lambda_{\bar{z}}-i\A_\sigma \eta
\right]
\right|_{\text{at fixed point}}
= -\frac{i\beta\e k}{2}(\sign(\sigma^*)-1).
\ee

Combining the above, we can evaluate the partition function
as a summation over
the fixed points, which are solutions $\sigma^*$ to $U(\sigma)=0$
\be
\begin{split}
Z
&= \sum_{\sigma^*}\prod_{n=-\infty}^{\infty}
\frac{1}{\omega_n+\e}
\frac{i\omega_n+G^{-1}U'(\sigma^*)}{\sqrt{\omega_n^2+(G^{-1}U'(\sigma^*))^2}}
e^{-\frac{i\beta\e k}{2}(\sign(\sigma^*)-1)}\\
&= \frac{\beta}{2i\sin \frac{\beta \e}{2}} \sum_{\sigma^*} \sign(U'(\sigma^*))
e^{-\frac{i\beta\e k}{2}(\sign(\sigma^*)-1)},
\end{split}
\ee
where $\omega_n=\frac{2\pi n}{\beta}$ is eigenvalue of the operator $-i\del_\tau$
with the periodic boundary condition.
Using the monopole potential of $U(\sigma) = \frac{k}{2|\sigma|} -\zeta$,
we have two fixed points $\sigma^*=\pm\frac{k}{2\zeta}$,
so we obtain the index explicitly
\be
Z =\frac{\beta y^k}{y-y^{-1}}(y^k - y^{-k}),
\ee
where $y \equiv e^{\frac{i\beta\e}{2}}$. This is nothing but the refined index of the monopole with the charge
$\gamma_m=(0,k)$ and the electron with the charge $\gamma_e =(1,0)$.
Taking the limit $\e \to 0 $, we obtain the number of the bound states (index) up to an
irrelevant overall contant.
We also would like to note here that the result of the localization formula does not depend on
the conformal factor $G(\vec{X})$ of the conformally flat metric.
The fixed points and the index is determined only by the data of the D-term potential $U(\vec{X})$
at the critical points.
This explains why the wall crossing formula is also valid for the BPS bound states
in the gauge theory without the gravity (in the flat background).

For the multi-particle case, we can generalize the localization formula to
\be
\begin{split}
Z&=\left(\frac{\beta}{y-y^{-1}}\right)^{n}
\sum_{\sigma^*:U_I(\{\sigma^*\})=0} \prod_I \sign(U'_I(\{\sigma^*\}))
y^{\sum_{I<J}
k_{IJ}\sign(\sigma_I^*-\sigma_J^*) }
.
\end{split}
\ee
This is the building block of the formula for the Abelian nodes derived in \cite{Manschot:2010qz, Manschot:2011xc} up to the overall constant (phase factor)\footnote{
Since the diagonal $U(1)$ factor (the center of motion) is decoupled in this derivation,
we should replace $n$ to $n-1$ to compare with the MPS formula.
}.
We however can not drive the localization formula for the non-Abelian nodes,
since we are considering only the non-linear sigma model with the Abelian gauge groups $G=U(1)^n$.
To obtain the localization formula in the Coulomb branch for the non-Abelian node, we have to
go back to the original gauged linear sigma model.
The quiver gauge theory contains effects from off-diagonal components of the non-Aberian gauge group,
which represent interactions among the identical superparticles.
In the following section, we will derive the localization formula of the quiver quantum mechanics from the point of view of
the gauged linear sigma model,
after integrating out all of massive Higgs fields.

\section{Gauged Linear Sigma Model Approach} \label{sec:glsm}

From this section, we discuss the gauged linear sigma model and show how to evaluate the refined index in the Coulomb branch by using the localization. As well as the non-linear sigma model case, we remove the center of mass part and consider the relative part only.
After decoupling the overall $U(1)$ part, the theory becomes the $\Ncal=4$ $U(N)$ supersymmetric quantum mechanics with $k$ fundamental chiral multiplets. The construction of the model is reviewed in the Appendix \ref{appconventions}. We basically
follow the conventions used in \cite{Ohta:2014ria}.

The BRST transformations, which is a part of the supersymmetry, for a vector multitplet are given by
\be
\begin{array}{lcl}
QZ=i\lambda_{z}, &&Q\lambda_{z}=i(\Dcal_{\tau}Z+[\sigma ,Z]+i\e Z), \\
Q\Zbar=-i\lambda_{\zbar}, && Q\lambda_{\zbar}=-i(\Dcal_{\tau}\Zbar +[\sigma,\Zbar]-i\e \Zbar),\\
QA=i\eta, &&\\
Q\sigma=\eta, && Q\eta=-\Dcal_{\tau}\sigma,\\
QY_{\R}=i(\Dcal_{\tau} \chi_{\R} +[\sigma, \chi_{\R}]), && Q\chi_{\R}=iY_{\R}, \label{eq:brstvec}
\end{array}
\ee
and 
those  for $k$ fundamental chiral multiplets are
\be
\begin{array}{lcl}
Qq_a =i\psi_a,&& Q\psi_a=i(\Dcal_{\tau} q_a+\sigma q_a +i\e_a q_a),\\
Q\qbar_a =-i\psibar_a, && Q\psibar_a=-i(\Dcal_{\tau}\qbar_a-\qbar_a \sigma -i \e_a \qbar_a),\\
QY_{\C,a} =i(\Dcal_{\tau}\chi_{\C,a}+\sigma \chi_{\C,a}+i(\e+\e_a)\chi_{\C, a}), && Q\chi_{\C,a}=iY_{\C,a},\\
Q\Ybar_{\C,a} =i(\Dcal_{\tau}\chibar_{\C,a}-\chibar_{\C,a}\sigma-i(\e+\e_a)\chibar_{\C, a}), && Q\chibar_{\C,a}=i\Ybar_{\C,a},
\end{array}
\ee
where $\Dcal_\tau \equiv \del_\tau +iA\cdot$ and {\it dot} denotes the action to the representation for each field.
Using this transformation law, the action takes the following $Q$-exact form:
\begin{align}
S&=S_{V}+S_C, \\
S_{V}&=\frac{1}{2g^2}\int d\tau \Tr \bigg[\frac{1}{2}Q(\lambda_z \overline{Q\lambda_z})+\frac{1}{2}Q(\lambda_{\zbar}\overline{Q\lambda_{\zbar}})+Q(\eta \overline{Q\eta})-Q(\chi_{\R} \overline{Q\chi_{\R}})+2iQ(\chi_{\R}\mu_{\R}^V)\bigg],  \label{eq:vectoraction} \\
S_C&=\int d\tau \Tr \bigg[\sum_{a=1}^k\bigg(\frac{1}{2}Q(\psi_a\overline{Q\psi_a})+\frac{1}{2}Q(\psibar_a \overline{Q\psibar_a})
-\frac{1}{2}Q(\chi_{\C,a}\overline{Q\chi_{\C,a}})-\frac{1}{2}Q(\chibar_{\C,a} \overline{Q\chibar_{\C,a}}) \notag \\
&-iQ(\chibar_{\C,a} \mu_{\C,a})-iQ(\chi_{\C,a}\mubar_{\C,a})\bigg)+2iQ(\chi_{\R}\mu_{\R}^C)\bigg], \label{eq:chiralaction}
\end{align}
where
\begin{align}
\mu_{\R}^V&=\frac{1}{2}[Z,\Zbar], \\
\mu_{\R}^C&=\frac{1}{2}\bigg(\sum_{a=1}^kq_a\qbar_a -\zeta\bigg), \\
\mu_{\C,a}&=Zq_a,~~~~~~~~~~\mubar_{\C,a}=\qbar_a\Zbar,
\end{align}
are moment maps corresponding to the D and F term conditions.
Since the action is written in $Q$-exact form, we can expect that the refined index should be  independent of the gauge coupling $g$.
So we can use the weak coupling approximation ($g\to 0$ limit) exactly in order to evaluate the refined index.

\subsection{Localization} \label{sec:derivation}

At first, let us see the gauge fixing. We impose the following gauge conditions:
\begin{align}
A_{\alpha}=0,~~~~~\del_\tau A_I=0,
\end{align}
where $A=A_IH^I+A_{\alpha}E^{\alpha}$ $(I=1,\cdots, N)$ and $H^I, E^{\alpha}$ are the Cartan and non-Cartan generators of $U(N)$, respectively. The constant modes of $A_I$   remain unfixed and we denote them by $\alpha_I$.
Then,  the Fadeev-Popov determinant becomes
\be
\begin{split}
\Delta_{\mathrm{FP}}&=\prod_{I\neq J}^N\prod_{n=-\infty}^\infty(\omega_n+\alpha_I-\alpha_J).
\end{split}
\ee
These $\alpha_I$ represent the degrees of freedom for  the Wilson loop along $\tau$-direction. Since the Wilson loop is the  gauge invariant quantity, we have to take into account them.

By taking $g\to 0$ limit, $S_V$ becomes dominant compared to $S_C$. We expand the fields for the vector multiplet as follows:
\begin{align}
Z&=0+g\tilde{Z}, ~~~~~~~~~~~~~~~\lambda_{z}=0+g\tilde{\lambda}_{z},  \notag \\
\sigma&=\sigma_0+g\tilde{\sigma}, ~~~~~~~~~~~~~~\eta=\eta_0+g\tilde{\eta},   \\
Y_{\R}&=Y_{\R,0}+g\tilde{Y}_{\R}, ~~~~~~~~~\chi_{\R}=\chi_{\R,0}+g\tilde{\chi}_{\R}, \notag
\end{align}
where $\sigma_0, \eta_0, Y_{\R,0}, \chi_{\R,0}$ are  constant diagonal matrices and the fields with tilde are  canonically normalized. The moduli of the theory are given by $\alpha, \sigma_0, \eta_0, \chi_{\R,0}$. Although $Y_{\R,0}$ is not the moduli, it plays an important role later \cite{Benini:2013nda,Benini:2013xpa,Hwang:2014uwa,Hori:2014tda}.

To obtain the 1-loop determinant for the vector multiplet, it is convenient to define a real supervector,
\begin{align}
\Vt=(\sigmatilde, \Yt_{\R},\etat, \chit_{\R}),
\end{align}
and a complex supervector,
\begin{align}
\tilde{W}=(\tilde{Z}, \tilde{\lambda}_z).
\end{align}
In the $g\to 0$ limit, (\ref{eq:vectoraction}) becomes\footnote{For the  auxiliary fields, we  change $Y_{\R}\to iY_{\R}, Y_{\C}\to iY_{\C}, \Ybar_{\C}\to i\Ybar_{\C}$ to avoid  divergent Gaussian integrations.}
\begin{align}
S_{V}=\frac{\beta}{2g^{2}}\Tr \big(Y_{\R,0}^2\big)+\frac{1}{2}\int d\tau \Tr \big(\bar{\Vt} M_{V}\Vt+\bar{\tilde{W}}M_{W}\tilde{W}\big),
\end{align}
where
\begin{align}
M_{V}&=\left(\begin{array}{cccc}-\Dcal_\tau^2 & 0 & -ad(\eta_0) & -ad(\chi_{\R,0}) \\0 & 1 & 0 & 0 \\-ad(\eta_0) & 0 & \Dcal_\tau+ad(\sigma_0) & 0 \\-ad(\chi_{\R,0}) & 0 & 0 & \Dcal_\tau +ad(\sigma_0)\end{array}\right), \\
M_{W}&=\left(\begin{array}{cc}-\Dcal_\tau^2+(ad(\sigma_0))^2-2i\e \Dcal_\tau +\e^2 -iad(Y_{\R,0}) & \sqrt{2}iad(\lambda_{1,0}) \\ -\sqrt{2}iad(\lambdabar_{1,0}) & \Dcal_\tau -ad(\sigma_0)+i\e \end{array}\right),
\end{align}
are the  supermatrices, $ad(X)Y\equiv [X,Y]$ and $\lambda_1$ is the first component of $\lambda_{\alpha}$ (See (\ref{eq:deffermion})).
From now on, we suppress the subscript $0$ for the constant modes.

The 1-loop determinant for the vector multiplet is given by
\begin{align}
\Delta_V=\Delta_{\mathrm{FP}}\frac{1}{\sqrt{\mathrm{Sdet} M_{V}}}\frac{1}{\mathrm{Sdet}M_W},
\end{align}
where $\mathrm{Sdet}$ denotes the superdeterminant. 
So we find
\begin{align}
\Delta_V&=(-1)^{N^2}\bigg(\frac{\beta}{y-y^{-1}}\bigg)^N\prod_{n=-\infty}^\infty\prod_{I\neq J}\frac{(\omega_n+\alpha_{IJ}+\e +i\sigma_{IJ})(\omega_n +\alpha_{IJ} -i\sigma_{IJ})}{|\omega_n+\alpha_{IJ}+\e-i\sigma_{IJ}|^2  -iY_{\R,IJ}} \notag \\
& \exp \bigg[\sum_{n=-\infty}^\infty\sum_{I\neq J} \frac{2i\lambdabar_{1,IJ}\lambda_{1,IJ}}{(\omega_n+\alpha_{IJ}+\e +i\sigma_{IJ})((\omega_n+\alpha_{IJ}+\e)^2+\sigma_{IJ}^2  -iY_{\R,IJ})}\bigg]. \label{eq:vector1loop} 
 \end{align}

Next, we consider the 1-loop determinant for chiral multiplets. In the $g\to 0$ limit, (\ref{eq:chiralaction}) becomes
\begin{align}
S_C&= i\beta \zeta \Tr Y_{\R} +\sum_{a=1}^k\int d\tau \Tr \big(\Xibar_{q,a} M_{q,a} \Xi_{q,a}+\Xibar_{Y,a} M_{Y,a} \Xi_{Y,a}\big),
\end{align}
where
\begin{align}
\Xi_{q,a}&=(q_a, \psi_a), \\
\Xi_{Y,a}&=(Y_{\C,a},\chi_{\C,a}),
\end{align}
are the complex supervectors and
\begin{align}
M_{q,a}&=\left(\begin{array}{cc}-\Dcal_\tau^2+\sigma^2-2i\e_a \Dcal_\tau +\e_a^2-iY_{\R}  & \sqrt{2}i\lambda_{1}  \\-\sqrt{2}i\lambdabar_{1}  & \Dcal_\tau-\sigma+i\e_a \end{array}\right), \\
M_{Y,a}&=\left(\begin{array}{cc} 1 & 0   \\ 0& \Dcal_\tau+\sigma+i(\e+\e_a)\end{array}\right),
\end{align}
are the supermatrices. 
Since the 1-loop determinant for the chiral multiplets is given by
\begin{align}
\Delta_C&=\prod_{a=1}^k\frac{1}{\mathrm{Sdet}M_{q,a}}\frac{1}{\mathrm{Sdet}M_{Y,a}},
\end{align}
we find
\begin{align}
\Delta_C&=(-1)^{kN}\prod_{a=1}^k\prod_{I=1}^N\prod_{n=-\infty}^\infty\frac{(\omega_n+\alpha_I+i\sigma_I +\e_a)(\omega_n+\alpha_I -i\sigma_I +\e +\e_a)}{|\omega_n +\alpha_I -i\sigma_I +\e_a|^2-iY_{\R,I}}  \notag\\ 
& \exp \bigg[\sum_{a=1}^k\sum_{I=1}^N\sum_{n=-\infty}^\infty \frac{2i\lambdabar_{1,I}\lambda_{1,I}}{((\omega_n+\alpha_I+\e_a)^2+\sigma_I^2-iY_{\R,I})(\omega_n+\alpha_I+i\sigma_I+\e_a)}\bigg]. \label{eq:chiral1loop}
\end{align}

After the Gaussian integrations, 
the refined index becomes
\begin{align}
\Zcal_N&=\frac{1}{N!}\int \prod_{I=1}^N \frac{d\alpha_I}{2\pi} \frac{d\sigma_I}{\sqrt{2\pi}}  \frac{dY_{\R,I}}{\sqrt{2\pi}} d\lambda_{1,I} d\lambdabar_{1,I}\Delta_V\Delta_C\exp\bigg[-\sum_{I=1}^N\bigg(\frac{\beta}{2g^2}Y_{\R,I}^2+i\beta \zeta Y_{\R,I}\bigg)\bigg], \label{eq:nonabelrefind}
\end{align}
 where $1/N!$ comes from the Weyl permutation. Since the refined index is periodic under $\alpha_I \to \alpha_I+\frac{2\pi}{\beta}$, we  integrate over $\alpha_I$ only in the fundamental region. This expression itself has  already been obtained in \cite{Hwang:2014uwa,Hori:2014tda}, but from this expression we can show that the moduli are localized at the Coulomb branch fixed points by taking $\zeta/\beta \to 0$ limit.

\subsection{Abelian case}
Firstly, we consider the Abelian case. 
In this case, the refined index becomes
\begin{align}
\Zcal_1&=\int  \frac{d\alpha}{2\pi} \frac{d\sigma}{\sqrt{2\pi}}  \frac{dY_{\R}}{\sqrt{2\pi}} d\lambda_{1} d\lambdabar_{1}\Delta_V\Delta_C\exp\bigg[-\frac{\beta}{2g^2}Y_{\R}^2-i\beta \zeta Y_{\R}\bigg], \label{eq:z1int}
\end{align}
where
\begin{align}
\Delta_V&=\frac{-\beta}{y-y^{-1}}, \label{DeltaV1} \\
\Delta_C&=(-1)^k\prod_{a=1}^k\prod_{n=-\infty}^{\infty}\frac{(\omega_n +\alpha +\e_a +i\sigma)(\omega_n +\alpha +\e +\e_a -i\sigma)}{(\omega_n+\alpha+\e_a)^2+\sigma^2-iY_{\R}} \notag \\
& \exp\bigg[\sum_{a=1}^k\sum_{n=-\infty}^\infty \frac{2i \lambdabar_1\lambda_1}{(\omega_n+\alpha+\e_a+i\sigma)((\omega_n+\alpha+\e_a)^2+\sigma^2-iY_{\R})}\bigg]. \label{DeltaC1}
\end{align}
In the $g\to 0$ limit, we can expand (\ref{DeltaC1}) around $Y_{\R}=0$. After integrating over $\lambda_1, \lambdabar_1$,  we find
\begin{align}
\Delta_C &=\bigg[\sum_{a=1}^k\sum_{n=-\infty}^\infty \frac{2i }{(\omega_n+\alpha+\e_a+i\sigma)((\omega_n+\alpha+\e_a)^2+\sigma^2)} + \Ocal(Y_{\R})\bigg] \notag \\
&\prod_{a=1}^{k}\frac{\sinh \frac{\beta}{2}(\sigma +i (\e+\e_a +\alpha))}{\sinh \frac{\beta}{2}(\sigma +i (\e_a +\alpha))} \exp \bigg[i\beta Y_{\R}\sum_{a=1}^k \frac{\sinh \beta \sigma}{2\sigma (\cosh \beta \sigma -\cos \beta (\alpha+\e_a))}+\Ocal(Y_{\R}^2)\bigg]. \label{eq:chiral1loop1}
\end{align}
The first term in the exponential is for a 1-loop correction of the FI term \cite{Denef:2002ru}, so the effective FI parameter is given by
\begin{align}
\zeta_{\mathrm{eff}}(\alpha,\sigma)=\zeta-\zeta_{\mathrm{1-loop}}(\alpha,\sigma),
\end{align}
where
\begin{align}
\zeta_{\mathrm{1-loop}}(\alpha,\sigma)&=\sum_{a=1}^k \frac{\sinh \beta \sigma}{2\sigma (\cosh \beta \sigma -\cos \beta (\alpha+\e_a))}.
\end{align}
The function $\zeta_{\mathrm{1-loop}}(\alpha,\sigma)$ is always positive  and has the following asymptotics:
\begin{align}
\zeta_{\mathrm{1-loop}}(\alpha,\sigma)&\stackrel{\beta \sigma \to \infty}{\simeq}\frac{k}{2|\sigma|}, \\
\zeta_{\mathrm{1-loop}}(\alpha,\sigma)&\stackrel{\beta \sigma \to 0}{\simeq} \frac{\beta}{4}\sum_{a=1}^k \frac{1}{\sin^2 \frac{\beta}{2}(\alpha+\e_a)}.
\end{align}
We note that there is no more quantum correction for the FI parameter due to the four supersymmetries \cite{Denef:2002ru}. 

Integrating over $Y_{\R}$, the following factor appears in the refined index:
\begin{align}
\exp\bigg[ -\frac{\beta g^2}{2}\zeta_{\mathrm{eff}}(\alpha,\sigma)^2\bigg]. \label{eq:expfactor}
\end{align}
So if we assume 
\begin{align}
 \beta g^2 \zeta^2 \gg 1, \label{eq:finallocal}
\end{align}
 the moduli are localized at
 \begin{align}
  \zeta_{\mathrm{eff}}(\alpha,\sigma)=0. \label{eq:effD}
 \end{align}
 This is the effective D-term condition. If $\zeta <0$, this condition can not be satisfied and  the refined index becomes zero.  This is
 nothing but the wall crossing phenomenon \cite{Denef:2007vg}. 

Let us assume $\zeta >0$.
If we take $\zeta/\beta \to 0$, we find that $\sigma$ is localized at
\begin{align}
\sigma^{\ast}=\pm \frac{k}{2\zeta}, \label{fixedCoulomb}
\end{align}
which are the fixed points for the Coulomb branch. On the other hand, if we take $\zeta/\beta \to \infty$, we find that the moduli are localized at 
\begin{align}
\sigma^\ast=0,~~~~~~\alpha^{\ast}=-\e_a, \label{eq:higgsfixedpt}
\end{align}
which are the fixed points for the Higgs branch \cite{Ohta:2014ria}. After the residue integrations around  the fixed points (\ref{eq:higgsfixedpt}), we obtain  the Poincar\'{e} polynomial of $\C P^{k-1}$, which is the Higgs branch moduli space \cite{Cordova:2014oxa,Hori:2014tda,Ohta:2014ria}. Now, we show that the same result can be derived by taking the Coulomb branch limit $\zeta/\beta \to 0$. 

Here, we give a comment. After integrating out the chiral multiplets, the quadratic term for $Y_{\R}$ in the effective action is given by
\begin{align}
-\frac{\beta}{2g^2}\bigg(1+\frac{g^2 k}{4|\sigma|^3}\bigg)Y_{\R}^2,
\end{align}
where the last term comes from  $\Ocal(Y_{\R}^2)$ term of (\ref{eq:chiral1loop1}) in the Coulomb branch limit. From the viewpoint of the nonlinear sigma model discussed in section \ref{sec:nlsm}, the coefficient of $Y_{\R}^2$ is just the metric on the moduli space. But in our analysis, we have neglected the last term. Using (\ref{fixedCoulomb}), the condition for the last term to be much  smaller than the first term is  
\begin{align}
g^2\zeta^3 \ll 1. \label{eq:weak}
\end{align} 
This is the case where the metric on the moduli space is almost flat so that the interpretation as  massless  closed string exchange is possible \cite{Douglas:1996yp}. 

To summarize, our Coulomb branch analysis is valid when
\begin{align}
1 \ll \beta g^2 \zeta ^2 \ll \frac{\beta}{\zeta}.
\end{align}
This inequality is satisfied if we take $\beta \to \infty$ with  $g$ and  $\zeta$ fixed. Of course, the refined index does not change with this limit because it does not depend on $\beta$ \cite{Witten:1982df}.

In the Coulomb branch limit, (\ref{eq:chiral1loop1}) becomes
\begin{align}
\Delta_C \simeq \frac{\beta k \sigma}{2|\sigma|^3} e^{\frac{ik\beta \e}{2}\sign(\sigma)}\exp\bigg[\frac{i\beta kY_{\R}}{2|\sigma|}\bigg]. \label{eq:simplification}
\end{align}
We note that the dependence on $\alpha$ and $\e_a$ in $\Delta_C$ disappears, so the integration over $\alpha$ becomes trivial.  After the Gaussian integration over $\sigma$ around the fixed points, we find
\begin{align}
\Zcal_1&=\frac{(-1)^{k+1}}{y-y^{-1}}\sum_{\sigma^{\ast}=\pm \frac{k}{2\zeta}}\sign(\sigma^{\ast})y^{k\cdot \sign(\sigma^{\ast})}. \label{eq:res1}
\end{align}
Therefore, the refined index is written as a summation over the Coulomb branch fixed points (\ref{fixedCoulomb}).

Compared to the MPS formula which is given in the Appendix \ref{sec:MPS}, we find that (\ref{eq:res1}) takes the same form as (\ref{eq:mpsg}). Summing over the fixed points, we obtain
\begin{align}
\Zcal_1=(-1)^{k+1}\frac{y^k-y^{-k}}{y-y^{-1}}, \label{eq:cp}
\end{align}
which agrees with (\ref{eq:MPS1})  when we set 
\begin{align}
 \gamma_{12}&=k, \label{eq:gamma12k} \\
\Omega^S(M,N)&=
\begin{cases}
&1~~~~~~~\text{for}~(M,N)=(1,0), (0,1) \\
&0~~~~~~~\text{otherwise}. 
\end{cases} \label{eqpmegas}
\end{align}
Our result also agrees with the Poincar\'{e} polynomial of $\mathbb{C}P^{k-1}$.

It is interesting that in contrast to the Higgs branch localization, there is no residue integration over the moduli in the Coulomb branch localization. But if we consider the non-Abelian case, we encounter  some residue integrations over the gauge fields.

\subsection{Non-Abelian case}
Next, we consider the non-Abelian case. Expanding the 1-loop determinants around $Y_{\R,I}=0$ with $g\to 0$ limit, the refined index becomes
\begin{align}
\Zcal_N&\simeq \frac{(-1)^{N^2+kN}}{N!}\bigg(\frac{1}{y-y^{-1}}\bigg)^N\int \prod_{I=1}^N \frac{\beta d\alpha_I}{2\pi}\frac{d\sigma_I}{\sqrt{2\pi}}\frac{dY_{\R,I}}{\sqrt{2\pi}} \notag \\
&\prod_{I=1}^N\bigg[\sum_{a=1}^k\sum_{n=-\infty}^\infty \frac{2i}{((\omega_n+\alpha_I+\e_a)^2+\sigma_I^2)(\omega_n+\alpha_I+i\sigma_I+\e_a)}\bigg] \notag \\
&\prod_{I\neq J}^N\frac{\sin \frac{\beta}{2}(\alpha_{IJ}-i\sigma_{IJ})}{\sin \frac{\beta}{2}(\alpha_{IJ}-i\sigma_{IJ}+\e)}\prod_{a=1}^k\prod_{I=1}^N\frac{\sinh \frac{\beta}{2}(\sigma_I +i(\alpha_I+\e+\e_a))}{\sinh \frac{\beta}{2}(\sigma_I +i(\alpha_I+\e_a))} \notag \\
&\prod_{I=1}^N\exp\bigg(-\frac{\beta}{2g^2}Y_{\R,I}^2-i\beta \zeta_{\mathrm{eff},I} Y_{\R,I}\bigg),
\end{align}
where 
\begin{align}
\zeta_{\mathrm{eff},I}=\zeta-\sum_{a=1}^k\frac{\sinh \beta \sigma_I}{2\sigma_I (\cosh \beta \sigma_I -\cos \beta (\alpha_I +\e_a))}.
\end{align}
There is no 1-loop contribution to the FI parameter from the vector multiplet in the Coulomb branch limit. 
The effective D-term conditions are given by $\zeta_{\mathrm{eff},I}=0$. Taking $\beta \to \infty$ limit, this condition becomes
\begin{align}
|\sigma_I|=\frac{k}{2\zeta}.
\end{align}
So if $\zeta >0$,  $\sigma_I$ are localized at $\sigma_I^\ast=\pm \frac{k}{2\zeta}$. In what follows, we assume $\zeta > 0$.

In the $\beta \to \infty$ limit, the 1-loop determinant for the chiral multiplets can be simplified as in (\ref{eq:simplification}) because $\sigma_I^{\ast} \neq 0$. But this kind of simplification does not always occur for the 1-loop determinant of the vector multiplet because $\sigma^{\ast}_{IJ}$ can be zero. Moreover, when $\sigma_{IJ}=0$, there are poles at $\alpha_{I}=\alpha_J \pm \e$ which are just on the contours of $\alpha_I$.

To avoid this, we only simplify $\Delta_C$ and start from the following expression:
\begin{align}
\Zcal_N&=\frac{(-1)^{N^2+kN}}{N!}\bigg(\frac{1}{y-y^{-1}}\bigg)^N\int \prod_{I=1}^N \frac{\beta d\alpha_I}{2\pi}\frac{dY_{\R,I}}{\sqrt{2\pi}}\frac{d\sigma_I}{\sqrt{2\pi}} \prod_{I=1}^N \bigg(\frac{\beta k \sigma_I}{2|\sigma_I|^3}y^{k\cdot \sign(\sigma_I)}\bigg) \notag \\
&  \prod_{I=1}^N \exp \bigg[-\frac{\beta}{2g^2}Y_{\R,I}^2-i\beta Y_{\R,I}\bigg(\zeta -\frac{k}{2|\sigma_I|}\bigg)\bigg] \notag \\
& \prod_{n=-\infty}^\infty\prod_{I\neq J}^N\frac{(\omega_n+\alpha_{IJ}+\e +i\sigma_{IJ})(\omega_n +\alpha_{IJ} -i\sigma_{IJ})}{|\omega_n+\alpha_{IJ}+\e -i\sigma_{IJ}|^2  -iY_{\R,IJ}}. \label{eq:partitionfunc}
\end{align}
We expand $\sigma_I$ around the fixed points as
\begin{align}
\sigma_I=\sigma_I^{\ast}+\delta \sigma_I.
\end{align}
Then, we find
\begin{align}
\zeta -\frac{k}{2|\sigma_I|}&\simeq \sign(\sigma^\ast_I)\frac{2\zeta^2}{k}\delta \sigma_I,
\end{align}
and the second line of (\ref{eq:partitionfunc}) becomes
\begin{align}
\prod_{I=1}^N\exp \bigg[-\frac{\beta}{2g^2} \bigg(Y_{\R,I}+\sign(\sigma_I^{\ast})\frac{2ig^2\zeta^2}{k}\delta \sigma_I\bigg)^2\bigg]\prod_{I=1}^N\exp \bigg[-\frac{2\beta g^2 \zeta^4}{k^2}\delta \sigma_I^2\bigg]. 
\end{align}

Let us set
\begin{align}
\Yh_{\R,I}&= \frac{\beta^{1/2}}{g}\bigg(Y_{\R,I}+\sign(\sigma_I^{\ast})\frac{2ig^2\zeta^2}{k}\delta \sigma_I\bigg), \\
\delta \sigmah_I&=\frac{2\beta^{1/2}g\zeta^2}{k}\delta \sigma_I.
\end{align}
Then, (\ref{eq:partitionfunc}) becomes
\begin{align}
\Zcal_N&=\frac{(-1)^{N^2+kN}}{N!}\bigg(\frac{1}{y-y^{-1}}\bigg)^N\int \prod_{I=1}^N \frac{\beta d\alpha_I}{2\pi}\frac{d\Yh_{\R,I}}{\sqrt{2\pi}}\frac{d\delta \sigmah_I }{\sqrt{2\pi}}\notag \\
& \sum_{\{\sigma_I^{\ast}\}}\prod_{I=1}^N \sign(\sigma_I^{\ast})y^{k\cdot \sign(\sigma_I^{\ast})} \prod_{I=1}^N \exp \bigg[-\frac{1}{2}\Yh_{\R,I}^2\bigg]\exp \bigg[-\frac{1}{2}\delta \sigmah_I^2\bigg] \notag \\
& \prod_{n=-\infty}^\infty \prod_{I\neq J}^N\frac{(\omega_n+\alpha_{IJ}+\e +i(\sigma_{IJ}^{\ast}+c\delta \sigmah_{IJ}))(\omega_n +\alpha_{IJ} -i(\sigma_{IJ}^{\ast}+c\delta \sigmah_{IJ}))}{|\omega_n+\alpha_{IJ}+\e -i(\sigma_{IJ}^{\ast}+c\delta \sigmah_{IJ})|^2 -b(\sign(\sigma_I^{\ast})\delta \sigmah_{I}-\sign(\sigma_J^{\ast})\delta \sigmah_{J}) -ib\Yh_{\R,IJ}}, \label{eq:ZNgeneral}
\end{align}
where  $\{\sigma^\ast_I\}$ denotes all sets of the fixed points for $\sigma_I$ and
\begin{align}
b&=\frac{g}{\beta^{1/2}}, \label{eq: b} \\
c&=\frac{k}{2\beta^{1/2}g\zeta^2}. \label{eq: c}
\end{align}
In the following sections, we set $\zeta=1$ for simplicity.

\subsubsection{$N=2$} \label{sec:n2}

We would like to see some examples for fewer $N$ cases at the beginning.
Let us first consider the $N=2$ case. Up to the Weyl permutation, there are three sets of the fixed points for $\sigma_I$:
\begin{align}
(\sigma_1^{\ast}, \sigma_2^{\ast})=(+,-),~~~(+,+),~~~(-,-), \label{eq:n2patterns}
\end{align}
where we have only mentioned the signs of $\sigma^{\ast}_I$.  We define $\Zcal_N^{(N-j,j)}$ as a contribution to the refined index  when $N-j$ of  $\sigma_I^\ast$ are positive and $j$ of those are negative.

\begin{description}
\item[(i) $(\sigma_1^{\ast}, \sigma_2^{\ast})=(+,-)$]

This is the case when $\sigma_{12}^{\ast}\neq 0$,  so (\ref{eq:ZNgeneral}) can be simplified in the $\beta \to \infty$ limit. Then, we find 
\begin{align}
\Zcal_2^{(1,1)} &\stackrel{\beta \to \infty}{\simeq} \frac{1}{2}\frac{-1}{(y-y^{-1})^2}.
\end{align}
Including the case  when $(\sigma_1^{\ast}, \sigma_2^{\ast})=(-,+)$, the total contribution to the refined index of this type is given by $2\Zcal_2^{(1,1)}$.

\item[(ii) $(\sigma_1^{\ast}, \sigma_2^{\ast})=(+,+), (-,-)$]

These are the cases when $\sigma_{12}^{\ast}=0$, so we need a careful treatment for the   integrations. 
Let us consider the following integrals:
\begin{align}
 &\int \prod_{I=1}^2 \frac{\beta d\alpha_I}{2\pi} \frac{d\Yh_{\R,I}}{\sqrt{2\pi}} \frac{d\delta \sigmah_I}{\sqrt{2\pi}} \prod_{I=1}^2 \exp \bigg[-\frac{1}{2}\Yh_{\R,I}^2\bigg]\exp \bigg[-\frac{1}{2}\delta \sigmah_I^2\bigg] \notag \\
& \prod_{n=-\infty}^\infty \prod_{I\neq J}^2\frac{(\omega_n+\alpha_{IJ}+\e +ic\delta \sigmah_{IJ})(\omega_n +\alpha_{IJ} -ic\delta \sigmah_{IJ})}{|\omega_n+\alpha_{IJ}+\e -ic\delta \sigmah_{IJ}|^2  -\sign(\sigma_1^{\ast})b\delta \sigmah_{IJ}-ib\Yh_{\R,IJ}}. \label{eq:rank2pp}
\end{align}
Taking $g\to 0$ limit,  the second line of (\ref{eq:rank2pp}) becomes
\begin{align}
 \frac{\sin^2 \frac{\beta}{2}(\alpha_{12}-ic\delta \sigmah_{12})}{\sin \frac{\beta}{2}(\alpha_{12}+\e-ic \delta \sigmah_{12})\sin \frac{\beta}{2}(\alpha_{12}-\e-ic \delta \sigmah_{12})}. \label{eq:sinmeasure}
\end{align}
So there seems to be two poles for $\alpha_1$ at
\begin{align}
\alpha_1&=\alpha_2-\e+ic\delta \sigmah_{12}, \label{eq:polen21} \\
\alpha_1&=\alpha_2+\e+ic\delta \sigmah_{12}. \label{eq:polen22} 
\end{align}
But we will see that one of them is not  a pole, depending on the sign of $\delta \sigmah_{12}$. In fact, if we insert (\ref{eq:polen21}) or (\ref{eq:polen22}) into (\ref{eq:rank2pp}) before taking $g\to 0$ limit, we find that the second line of (\ref{eq:rank2pp}) includes the following factor in the $n=0$ mode:
\begin{align}
\frac{ik\beta \e}{\beta g^2} \frac{\delta \sigmah_{12}}{\sign(\sigma_1^\ast)\delta \sigmah_{12} +i \Yh_{\R, 12}}. \label{eq:zerofact}
\end{align}
Therefore, in the $\beta g^2\to \infty$ limit with  $\beta \e$ fixed,   (\ref{eq:zerofact}) vanishes.

To find which pole we should choose, we start with (\ref{eq:rank2pp}) and  deform the $\alpha_1$-contour    as in Figure \ref{fig:alphacont}. 
\begin{figure}
\begin{center}
\includegraphics[width=11cm,clip]{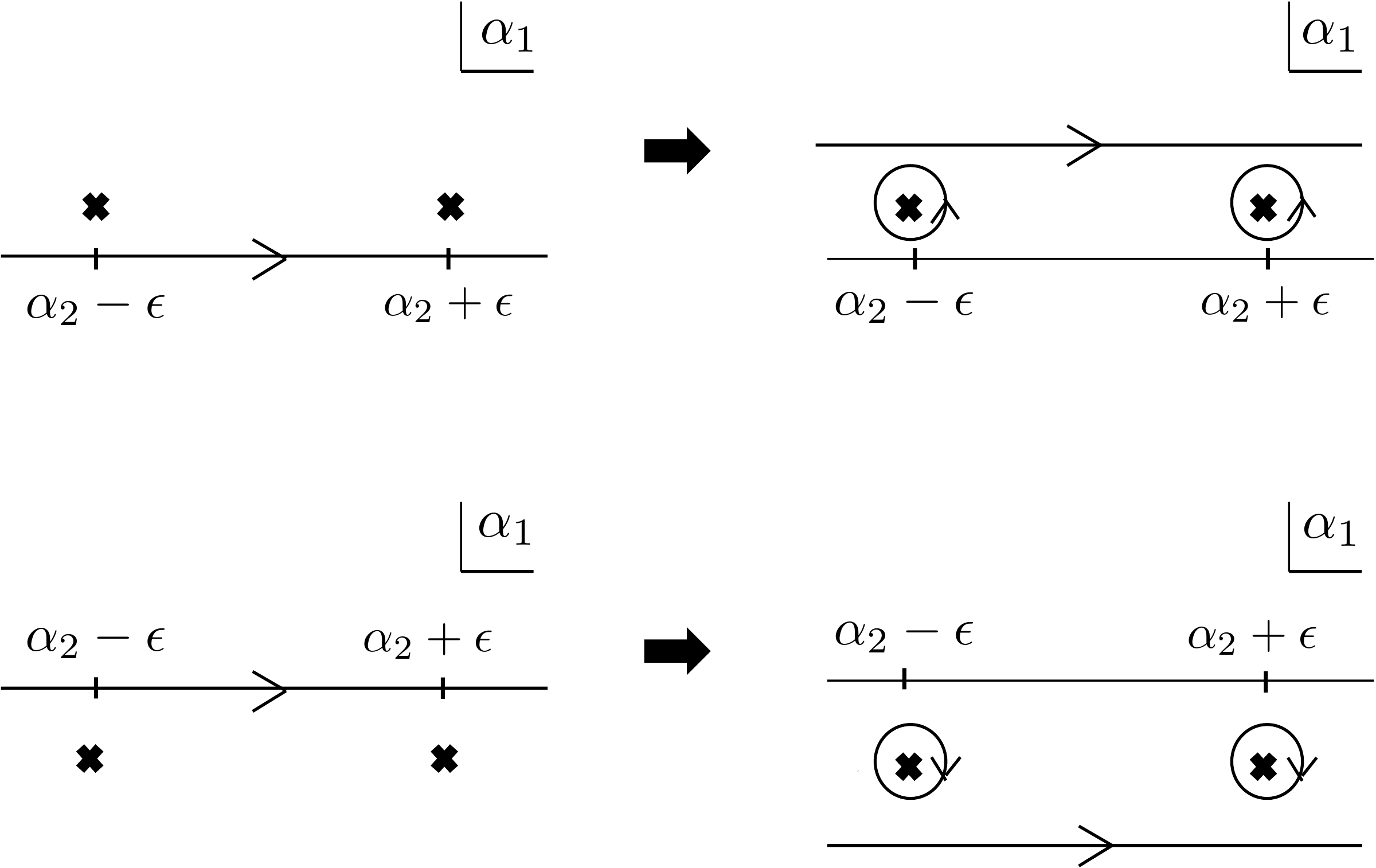}
\caption{The deformation of $\alpha_1$-contour. The original contour   lies on the real axis. If $\delta \sigmah_{12}>0$ ($\delta \sigmah_{12}<0$), we shift  the contour on the upper (lower) side. Then, the contour becomes a sum of two circular contours  around the poles  and a line contour with an imaginary part.}
\label{fig:alphacont}
\end{center}
\end{figure}
For the line contour with an imaginary part,  we can replace $\alpha_{12}\to \alpha_{12}\pm i\delta_{12}$ in the integrand.  This $\delta_{12}$ can be interpreted as a resolution for the degenerate fixed points: $\sigma_{12}^{\ast}=\pm \delta_{12}\neq 0$. Therefore, by taking $\beta\to \infty$, (\ref{eq:rank2pp}) becomes simplified as in the case of (i).

Next, we consider the contour integrals around (\ref{eq:polen21}) and (\ref{eq:polen22}). Let us see the former case.
In (\ref{eq:rank2pp}), we  have the following pole for $\Yh_{\R,1}$:
\begin{align}
\Yh_{\R,1}&=\Yh_{\R,2}-i\bigg(\frac{|\alpha_{12}+\e-i c \delta \sigmah_{12}|^2}{b} - \sign(\sigma_1^\ast)\delta \sigmah_{12}\bigg). \label{eq:Ypole1}
\end{align}
Setting $r$ as a radius of the contour, this expression becomes
\begin{align}
\Yh_{\R,1}&=\Yh_{\R,2}-i\bigg(\frac{r^2}{b} - \sign(\sigma_1^\ast)\delta \sigmah_{12}\bigg). \label{eq:Ypole3} 
\end{align}

 We assume  $r^2 > b |\delta \sigmah_{12}|$ and consider $r\to 0$ limit. If $\sign(\sigma_1^\ast )\delta \sigmah_{12}>0$,  the pole (\ref{eq:Ypole3}) collides with the $\Yh_{\R,1}$-contour  as $r$ approaches $0$ as in the left figure of  Figure \ref{fig:ycontour}. In this case, we evaluate (\ref{eq:rank2pp}) by taking  $g \to 0$ limit before $r\to 0$ limit. Then, the second line of (\ref{eq:rank2pp})  becomes (\ref{eq:sinmeasure}) and we evaluate the residue integral around the pole (\ref{eq:polen21}). 
 If $\sign(\sigma_1^\ast )\delta \sigmah_{12}<0$,  the pole (\ref{eq:Ypole3}) does not collide with the $\Yh_{\R,1}$-contour  even if $r$ becomes $0$ as in the right figure of  Figure \ref{fig:ycontour}. In this case, we take $r\to 0$ limit before $g\to 0$ limit and so (\ref{eq:rank2pp}) becomes 0 with $\beta g^2 \to \infty$. Using the same argument, we find that we take the pole (\ref{eq:polen22}) only when $\sign(\sigma_1^\ast )\delta \sigmah_{12} < 0$.

\begin{figure}
\begin{center}
\includegraphics[width=11cm,clip]{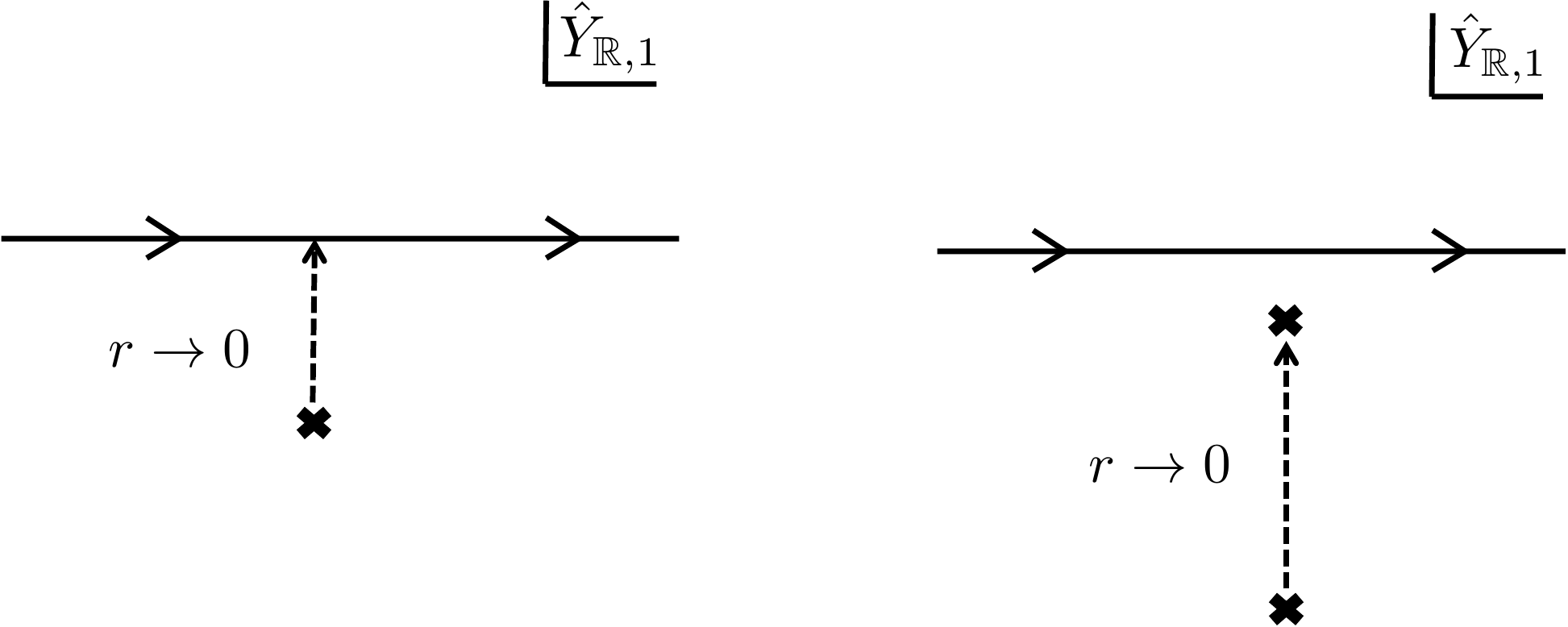}
\caption{A pole for $\Yh_{\R,1}$. The left hand side is for $\sign(\sigma_1^\ast )\delta \sigmah_{12} >0$ and the right hand side is for $\sign(\sigma_1^\ast )\delta \sigmah_{12} <0$}
\label{fig:ycontour}
\end{center}
\end{figure}

To summarize,  we choose the following pole for $\alpha_1$ in the $\beta \to \infty$ limit: 
\begin{align}
\alpha_1=
 \begin{cases}
 \alpha_2-\sign(\sigma_1^\ast)\e,~~~~~~~~(\text{with counterclockwise direction for}~\delta \sigmah_{12}>0), \\
 \alpha_2+\sign(\sigma_1^\ast)\e,~~~~~~~~(\text{with clockwise direction for}~\delta \sigmah_{12}<0).
 \end{cases}\label{eq:pole2finalpp}
\end{align}
The two contributions in (\ref{eq:pole2finalpp})  are related by the Weyl permutation and are  the same.

Therefore, we find
\begin{align}
\Zcal_2^{(2,0)}&=\frac{1}{2}\frac{y^{2k}}{(y-y^{-1})^2}\bigg[1-\frac{(y-y^{-1})^2}{y^2-y^{-2}}\bigg], \label{eq:220} \\
\Zcal_2^{(0,2)}&=\frac{1}{2}\frac{y^{-2k}}{(y-y^{-1})^2}\bigg[1+\frac{(y-y^{-1})^2}{y^2-y^{-2}}\bigg], \label{eq:202}
\end{align}
where the first terms come from the  $\alpha_1$-integral along the line contour with an imaginary part and the last terms come from the residue integral for $\alpha_1$.

\end{description}

Thus, the refined index for $N=2$  becomes
\begin{align}
\Zcal_2&=2\Zcal_2^{(1,1)}+\Zcal_2^{(2,0)}+\Zcal_2^{(0,2)} \notag \\
&=-\frac{y^{2k}-y^{-2k}}{2(y^2-y^{-2})}+\frac{1}{2}\bigg(\frac{y^{k}-y^{-k}}{y-y^{-1}}\bigg)^2, \label{eq:n2result}
\end{align}
which agrees with the MPS formula  (\ref{eq:MPS2}) term by term when we set (\ref{eq:gamma12k}) and (\ref{eqpmegas}). Carrying out the summation, we obtain
 \begin{align}
  \Zcal_2=\left[\begin{array}{c}k \\2\end{array}\right]_{-y},
 \end{align}
 where
\begin{align}
 \left[\begin{array}{c}k \\N\end{array}\right]_y&=\frac{[k]_y!}{[k-N]_y![N]_y!}, \\
 [x]_y!&=[x]_y[x-1]_y\cdots [1]_y.
\end{align}
Therefore, the refined index also agrees with   the Poincar\'{e} polynomial for the Grassmannian $Gr(2,k)$, which is the moduli space for the Higgs branch.
 
Let us give a comment. This is the case where the total charge of the BPS particles is given by $\gamma=\gamma_1+2\gamma_2\equiv (1,2)$, where $\gamma_{1}$ and  $\gamma_2$ are the primitive charge vectors. So the sum in the MPS formula (\ref{eq:MPSformula}) is composed of 
 \begin{align}
\gamma=\{(0,2)+(1,0), ~(0,1)+(0,1)+(1,0), ~(0,1)+(1,1)\}. \label{eq:n2contents}
 \end{align}
 The first one in (\ref{eq:n2contents}) corresponds to the case where two electrons are degenerate. The contribution of this type is given by the residue integral for the gauge fields in our Coulomb branch localization. 
 The second one  in (\ref{eq:n2contents}) corresponds to the case where there is no degeneracy for the BPS particles. It is clear that the case-(i) belongs to this type, but there is also the contribution to this type in the case-(ii), which is given by the  integral over the gauge field along a line contour with an imaginary part.
The last one in (\ref{eq:n2contents}) corresponds to the case where a monopole and an electron are degenerate. In our quiver quantum mechanics, this case does not occur.

\subsubsection{$N=3$}
Next, we consider $N=3$ case. Up to the Weyl permutation, there are four sets of the fixed points:
\begin{align}
(\sigma_1^{\ast}, \sigma_2^{\ast},\sigma_3^{\ast})=(+,+,-), ~(-,-,+), ~(+,+,+),~(-,-,-).
\end{align}

\begin{description}
\item[(i) $(\sigma_1^{\ast}, \sigma_2^{\ast},\sigma_3^{\ast})=(+,+,-),~(-,-,+)$]

Using the results of the $N=2$ case, we find
 \begin{align}
\Zcal_3^{(2,1)}
&=\frac{(-1)^{k+1}}{6}\frac{-y^{k}}{(y-y^{-1})^3}\bigg[1-\frac{(y-y^{-1})^2}{y^2-y^{-2}}\bigg], \\
\Zcal_3^{(1,2)}&=\frac{(-1)^{k+1}}{6}\frac{y^{-k}}{(y-y^{-1})^3}\bigg[1+\frac{(y-y^{-1})^2}{y^2-y^{-2}}\bigg].
\end{align} 
The total contribution to the refined index of this type is given by $3\Zcal_3^{(2,1)}+3\Zcal_3^{(1,2)}$.

\item[(ii) $(\sigma_1^{\ast}, \sigma_2^{\ast},\sigma_3^{\ast})=(+, +,+),~(-,-,-)$]

In each case, there is a situation where we have to choose two poles at the same time. 
From the discussion of section \ref{sec:n2}, the possible poles are the following:
\begin{align}
\alpha_{12}=
 \begin{cases}
 -\sign(\sigma_1^\ast)\e,~~~~~~~~(\text{with counterclockwise direction for}~\delta \sigmah_{12}>0), \\
 +\sign(\sigma_1^\ast)\e,~~~~~~~~(\text{with clockwise direction for}~\delta \sigmah_{12}<0),
 \end{cases} \label{eq:n3pole1}  \\
 \alpha_{13}=
 \begin{cases}
-\sign(\sigma_1^\ast)\e,~~~~~~~~(\text{with counterclockwise direction for}~\delta \sigmah_{13}>0), \\
+\sign(\sigma_1^\ast)\e,~~~~~~~~(\text{with clockwise direction for}~\delta \sigmah_{13}<0),
 \end{cases} \label{eq:n3pole2}  \\
 \alpha_{23}=
 \begin{cases}
-\sign(\sigma_1^\ast)\e,~~~~~~~~(\text{with counterclockwise direction for}~\delta \sigmah_{23}>0), \\
 +\sign(\sigma_1^\ast)\e,~~~~~~~~(\text{with clockwise direction for}~\delta \sigmah_{23}<0).
 \end{cases} \label{eq:n3pole3} 
 \end{align}
If we start from the $\alpha_1$-integral, the easiest way of the calculation is to focus on the pole  (\ref{eq:n3pole1})  and then double the result of the residue calculation because (\ref{eq:n3pole1}) and (\ref{eq:n3pole2}) are related by the Weyl permutation $2 \leftrightarrow 3$.  In the same way, for the next $\alpha_2$-integral, we only focus on  (\ref{eq:n3pole3}) for the residue calculation using the Weyl permutation. But if we pick up  two poles $\alpha_{12}=-\sign(\sigma_1^\ast)\e$ and $\alpha_{23}=\sign(\sigma_1^\ast)\e$ (or $\alpha_{12}=\sign(\sigma_1^\ast)\e$ and $\alpha_{23}=-\sign(\sigma_1^\ast)\e$), the contribution to the refined index is zero because the numerator in the integrand vanishes. So we find
\begin{align}
\Zcal_3^{(3,0)}&= \frac{(-1)^{k+1}}{6}\frac{y^{3k}}{(y-y^{-1})^3}\bigg[1-3\frac{(y-y^{-1})^2}{y^2-y^{-2}}+2\frac{(y-y^{-1})^3}{y^3-y^{-3}}\bigg], \\
\Zcal_3^{(0,3)}&=\frac{(-1)^{k+1}}{6}\frac{-y^{-3k}}{(y-y^{-1})^3}\bigg[1+3\frac{(y-y^{-1})^2}{y^2-y^{-2}}+2\frac{(y-y^{-1})^3}{y^3-y^{-3}}\bigg],
\end{align}
where the last terms are the contributions of when we pick up two poles.

\end{description}

Summing up  all these contributions, the refined index for $N=3$ becomes
\begin{align}
\Zcal_3&=\Zcal_3^{(3,0)}+3\Zcal_3^{(2,1)}+3\Zcal_3^{(1,2)}+\Zcal_3^{(0,3)} \notag \\
&=(-1)^{k+1}\bigg[\frac{y^{3k}-y^{-3k}}{3(y^3-y^{-3})}-\frac{(y^k-y^{-k})(y^{2k}-y^{-2k})}{2(y-y^{-1})(y^2-y^{-2})}+\frac{1}{6}\bigg(\frac{y^{k}-y^{-k}}{y-y^{-1}}\bigg)^3\bigg],
\end{align}
which agrees with the MPS formula (\ref{eq:MPS3}) term by term when we set (\ref{eq:gamma12k}) and (\ref{eqpmegas}). Also, this can be written as
\begin{align}
\Zcal_3=\left[\begin{array}{c}k \\3\end{array}\right]_{-y},
\end{align}
which agrees with the Poincar\'{e} polynomial for the Grassmannian $Gr(3,k)$.

\subsubsection{General $N$}
Finally, we would like to give the formula of the refined index for a general $N$.

At first, we  consider the case when $(\sigma_1^{\ast}, \cdots, \sigma_N^{\ast})=(\pm,\cdots,\pm)$. We define 
\begin{align}
I_N&=\int \prod_{I=1}^N \frac{\beta d\alpha_I}{2\pi}\prod_{I\neq J}^N\frac{\sin \frac{\beta}{2}(\alpha_I-\alpha_J)}{\sin\frac{\beta}{2}(\alpha_I-\alpha_J+\e)}, ~~~~~~~~(N \geq 2), \label{generalnint}
\end{align}
and denote it by $I_N^+$ if $(\sigma_1^{\ast}, \cdots, \sigma_N^{\ast})=(+,\cdots, +)$ and by $I_N^-$ if $(\sigma_1^{\ast}, \cdots, \sigma_N^{\ast})=(-,\cdots, -)$. Each $\alpha_I$-integral includes an integral along a line with an imaginary part and  residue integrals. For the residue integrals,   we  consider the poles at
\begin{align}
\alpha_I=\alpha_{I+1}-\sign(\sigma_1^{\ast})\e,~~~~~~(I=1,\cdots, N-1),
\end{align}
with counterclockwise direction, up to the Weyl permutations.

We define
\begin{align}
a_N^+&=(-1)^{N+1}\frac{(y-y^{-1})^N}{y^N-y^{-N}}, \\
a_N^-&=\frac{(y-y^{-1})^N}{y^N-y^{-N}}.
\end{align}
Integrating over $\alpha_1$, $I_N^{\pm}$ becomes
\begin{align}
I_N^\pm&=I_{N-1}^\pm \notag \\
&+ (N-1)\int \prod_{I=2}^N\frac{\beta d\alpha_I}{2\pi}i\beta\mathrm{Res}_{\alpha_1=\alpha_2\mp \e}\bigg( \prod_{I\neq J}^N\frac{\sin \frac{\beta}{2}(\alpha_I-\alpha_J)}{\sin\frac{\beta}{2}(\alpha_I-\alpha_J+\e)}\bigg). \label{eq:IN1}
\end{align}
Integrating over $\alpha_2$, (\ref{eq:IN1}) becomes
\begin{align}
I_N^{\pm}&=I_{N-1}^{\pm}+(N-1)\bigg[a_2^{\pm}I_{N-2}^{\pm} \notag \\
&+(N-2)\int \prod_{I=3}^N\frac{\beta d\alpha_I}{2\pi}(i\beta)^2\mathrm{Res}_{\substack{\alpha_1=\alpha_2\mp \e \\ \alpha_2=\alpha_3\mp \e}} \bigg(\prod_{I\neq J}^N\frac{\sin \frac{\beta}{2}(\alpha_I-\alpha_J)}{\sin\frac{\beta}{2}(\alpha_I-\alpha_J+\e)}\bigg)\bigg].
\end{align}
Continuing the integrations, we finally obtain the following recursion relation:
\begin{align}
 I_N^{\pm}&=\sum_{l=1}^{N}\frac{(N-1)!}{(l-1)!}a_{N-l+1}^{\pm}I_{l-1}^{\pm}, \label{eq:recursion}
\end{align}
where
\begin{align}
I_0^{\pm}= 1,~~~~~I_1^{\pm}=1.
\end{align}

Next, we consider the generic case. When $N-j$ of  $\sigma_I^{\ast}$ are positive and the others are  negative, there are  totally  ${k \choose N}$ patterns. So  the  contribution to the refined index is proportional to $ {k \choose N}I_{N-j}^+I_j^-$.

Therefore, the refined index for a generic $N$ is given by
\begin{align}
\Zcal_N&=(-1)^{N^2+kN}\bigg(\frac{1}{y-y^{-1}}\bigg)^N\sum_{j=0}^N(-1)^jy^{k(N-2j)}\frac{I_{N-j}^+}{(N-j)!}\frac{I_j^-}{j!}. \label{eq:generalref}
\end{align}
If we could solve (\ref{eq:recursion}) and insert the general solution of $I_N^\pm$ into (\ref{eq:generalref}), we would see the correspondence with the MPS formula for an arbitrary $N$. But it  seems to be too difficult. Instead, we have confirmed that (\ref{eq:generalref}) actually becomes ${k \brack N}_{-y}$ up to $N=4$, so our result will be correct in any $N$.

\section{Conclusion and Discussion}

In this paper, we have revisited the exact  analysis of the $\Ncal=4$ $U(1)\times U(N)$ supersymmetric quiver quantum mechanics and have shown how the Coulomb branch picture emerges in the localization calculation. We have discussed it from the viewpoint of both the nonlinear sigma model and gauged linear sigma model. 

We have seen that the localization
of the non-linear sigma model with the $\Omega$-background
(refined model)
is much easier than the original model that derives the Atiyah-Singer index theorem
on the Coulomb branch moduli,
since the flat directions are lifted up and the fixed points become isolated.
We also found that the localization does not depend on
the detail of the metric of the $\Ncal=4$ supersymmetric non-linear sigma model.
In particular, the conformal factor of the conformally flat metric is canceled out
in the evaluation of the 1-loop determinant.
The D-term potential and associated bulk gauge field crucially determine
the fixed point data, and
the factor and relative signs in the refined index. 

In the gauged linear sigma model approach,  we have found the effective D-term condition (\ref{eq:effD}) from the 1-loop determinant, which knows not only  the Higgs branch fixed points but also the Coulomb branch fixed points.   Using our Coulomb branch localization, the refined index can be written as a summation over the different sets of the Coulomb branch fixed points, which corresponds to the MPS formula. If the fixed points are not degenerate, the integrations over $\alpha_I$ in the refined index become trivial in the Coulomb branch limit $\zeta/\beta\to 0$ and the residue integrals do not appear in contrast to the Higgs branch localization \cite{Cordova:2014oxa,Hori:2014tda,Ohta:2014ria}. However, at the degenerate fixed points, we encounter the residue integrals over $\alpha_I$, which give the factors in the rational invariant (\ref{eq:rationalinv}).

It will be interesting to  generalize our analysis to  quiver  quantum mechanics with closed loops, which corresponds to the case where there exists scaling solutions on the supergravity side \cite{Denef:2007vg}. Recently, it has been announced that there are ``pure Higgs states" which can not be mapped to the multi-center solutions  in the supergravity \cite{Bena:2012hf,Lee:2012sc}. We would like to investigate   the pure Higgs states from the viewpoint of the Coulomb branch localization.

Our  localization technique might also be  useful to understand  Coulomb branches which are still unknown in some supersymmetric theories. For example, a D1-D5(-P) black hole is described by  a two-dimensional $\Ncal=(4,4)$ gauged linear sigma model and the Bekenstein-Hawking entropy has been derived from the Higgs branch analysis \cite{Strominger:1996sh,Callan:1996dv} (for review see \cite{David:2002wn}). On the other hand,  there  are infinitely many   supergravity solutions with the same charges as the black hole, which are known as the fuzzball solutions \cite{Lunin:2001fv,Lunin:2002iz,Lunin:2004uu,Giusto:2004id,Giusto:2004kj,Bena:2005va,Berglund:2005vb,Taylor:2005db,Bena:2006kb,Kanitscheider:2007wq}. The fuzzball solutions are conjectured to describe the microscopic geometries in the black hole \cite{Mathur:2005zp}, but the relation with the gauged linear sigma model is not known yet.  It will be interesting if we can find out  connections between the Coulomb branch fixed points of the gauged linear sigma model and the fuzzball solutions.

\section*{Acknowledgements}

We would like to thank
F.~Benini,
T.~Hollowood,
H.~Kanno,
P.~Kumar,
K.~Maruyoshi,
S.~Ramgoolam,
K.~Sakai,
N.~Sakai and
M.~Shigemori
for useful discussions and comments.
This work of KO is supported in part by JSPS KAKENHI Grant Number 14485514.


\section*{Appendix}

\appendix

\section{$\Ncal=4$ $U(N)$ supersymmetric quantum mechanics}\label{appconventions}

We here review an $\Ncal=4$ $U(N)$ supersymmetric quantum mechanics. This theory is obtained from the dimensional reduction of  four dimensional $\Ncal=1$ $U(N)$ supersymmetric gauge theory to one dimension and  possesses  $SU(2)_J\times U(1)_R$ global R-symmetries.

\subsection{Vector multiplet}

In one dimensional $\Ncal = 4$ supersymmetric theory, a vector multiplet is composed of a gauge field $A_0$, three real scalars $X_i$ $(i=1,2,3)$, four fermions $(\lambda_{\alpha},\bar{\lambda}_{\dot{\alpha}})$ $(\alpha,\dot{\alpha}=1,2)$, and an  auxiliary real scalar $D$.  All fields are in the adjoint representation of $U(N)$.
The representations of $SU(2)_J$ and   $U(1)_R$ charges of the vector multiplet are summarized in Table \ref{tab:rv}.
\begin{table}[h]
\centering
\begin{tabular}{c||c|c|c|c|c} 
~& $A_0$ & $X_i$ & $\lambda_{\alpha}$ & $\bar{\lambda}_{\dot{\alpha}}$ & $D$ \\ \hline
$SU(2)_J$ & $\mathbf{1}$ & $\mathbf{3}$ & $\mathbf{2}$ & $\bar{\mathbf{2}}$ & $\mathbf{1}$ \\ \hline
$U(1)_R$ & 0 & 0 & $1$ & $-1$ & 0 \\
\end{tabular} 
\caption{The R-symmetries of the vector multiplet.} 
\label{tab:rv}
\end{table}

Using these fields, the action is given by
\begin{align}
 S_V&=\frac{1}{g^2}\int dt \, \Tr\bigg[\frac{1}{2}(\Dcal_0X^i)^2+\frac{1}{4}[X^i,X^j]^2
 -i\lambdabar \sigmabar^0 \Dcal_0 \lambda +\lambdabar \sigmabar_i [X^i,\lambda]+\frac{1}{2}D^2 - g^2\zeta D\bigg],
 \label{vector action}
\end{align}
where $g$ is the gauge coupling, $\zeta$ is the FI parameter, and
\begin{align}
\Dcal_0\equiv \del_0+i[A_0,\cdot],
\end{align}
is a covariant derivative. The mass dimensions of $g$ and $\zeta$ are 
\begin{align}
 [g^2]&=3,~~~~~~~[\zeta]=-1.
\end{align}

The action (\ref{vector action}) is invariant under the following 
supersymmetric transformations:
\be
\begin{array}{lcl}
\delta A_{0} &=& -i\xi \sigma_{0} \lambdabar +i\lambda \sigma_0 \xibar, \\
\delta X^{i} &=& i\xi \sigma^{i} \lambdabar -i\lambda \sigma^i \xibar, \\
\delta \lambda &=& i\xi D +2 \sigma^{0i}\xi \Dcal_0 X_i+i\sigma^{ij}\xi [X_i,X_j] , \\
\delta D &=& -\xi \sigma^0 \Dcal_0 \lambdabar -i\xi \sigma^i [X_i, \lambdabar]
-\Dcal_0\lambda \sigma^0 \xibar -i[X_i,\lambda]\sigma^i \xibar,
\end{array}
\ee
where  $\xi_\alpha$ and $\bar{\xi}_{\dot{\alpha}}$ represent the supersymmetric parameters,
and in terms of the supercharges $Q_\alpha$, the supersymmetric variation is given by
\begin{align}
\delta=\xi Q+\xibar \Qbar.
\end{align}

We now introduce a linear combination of the supercharges by
\begin{align}
 Q&=\frac{i}{\sqrt{2}}(Q^1-\Qbar^{1})=\frac{i}{\sqrt{2}}(Q_2-\Qbar_{2}),
\end{align}
which will be called the BRST charge.
After the Wick rotation $t\to -i\tau$, we
 define linear combinations of the bosonic fields by
\be
\begin{array}{llll}
Z=X_1-iX_2, & \Zbar=X_1+iX_2,&
\sigma=X_3,& A= A_\tau,\\
Y_{\R}=D-\frac{1}{2}[Z,\Zbar],
\end{array}
\ee
and
\be
\begin{array}{lll}
\lambda_z=\sqrt{2}i\lambdabar_2, & \lambda_{\zbar}=-\sqrt{2}i\lambda_2, & \eta =-\frac{1}{\sqrt{2}}(\lambda_1+\lambdabar_1), \\
\chi_\R =\frac{i}{\sqrt{2}}(\lambda_1-\lambdabar_1),
\end{array} \label{eq:deffermion}
\ee
for the corresponding fermionic fields.
This operation is usually called  ``topological twisting''. The mass dimensions of the fields in the vector multiplet are
\begin{align}
[Z]&=[\Zbar]=[A]=[\sigma]=1, \notag \\
[\lambda_z]&=[\lambda_{\zbar}]=[\eta]=[\chi_{\R}]=\frac{3}{2}, \\
[Y_{\R}]&=2. \notag 
\end{align}

Under the introduced BRST symmetry, the fields are transformed as follws:
\be
\begin{array}{lcl}
QZ=i\lambda_{z}, &&Q\lambda_{z}=i(\Dcal_{\tau}Z+[\sigma ,Z]), \\
Q\Zbar=-i\lambda_{\zbar}, && Q\lambda_{\zbar}=-i(\Dcal_{\tau}\Zbar +[\sigma,\Zbar]),\\
QA=i\eta, &&\\
Q\sigma=\eta, && Q\eta=-\Dcal_{\tau}\sigma,\\
QY_{\R}=i(\Dcal_{\tau} \chi_{\R} +[\sigma, \chi_{\R}]), && Q\chi_{\R}=iY_{\R}.
\end{array}
\ee
The BRST transformations are nilpotent up to a translation along the time-direction
and (complexified) gauge transformation with the parameter $A+i\sigma$.

The Euclidean action of the theory (\ref{vector action}) is written as a $Q$-exact form:
\be
S_V=\frac{1}{2g^2}Q\int d\tau \Tr \bigg[\frac{1}{2}\lambda_{z} \overline{Q\lambda_{z}}+\frac{1}{2}\lambda_{\zbar}\overline{Q\lambda_{\zbar}} +\eta \overline{Q\eta} -\chi_{\R} \overline{Q\chi_{\R}} +2i\chi_{\R}\mu_{\R}\bigg],
\label{Q-exact action}
\ee
where $\mu_{\R}=\frac{1}{2}[Z,\Zbar]-g^2\zeta$ is a (real) moment map constraint which contains the original D-term constraint and describes the moduli space of the vacua.
After integrating out the auxiliary field $Y_\R$, we obtain the Euclidean action of the original matrix quantum mechanics.

The field redefinitions (topological twist) spoil the original R-symmetries,
but the theory is still invariant under the following twisted ``R-transformation" $U(1)_J'$, which acts on the fields by
\be
\begin{array}{lcl}
Z\to e^{i\theta_J}Z, && \lambda_z\to  e^{i\theta_J}\lambda_z,
\end{array}
\ee
with an R-transformation parameter $\theta_J$.\footnote{Using the $SU(2)_J$ transformation
\begin{align}
J_jX_i=-i\epsilon_{jik}X_k,~~~~~~~J^i \lambda=\sigma^{0i}\lambda \notag,
\end{align}
and the $U(1)_R$ transformation, we  find that the generator of  $U(1)'_J$ is given by $\frac{1}{2}(2J_3-R)$ \cite{Cordova:2014oxa,Hori:2014tda}. This generator commutes with the BRST charge $Q$, so we can define the refined index using this generator by
\begin{align}
I_{\mathit{ref}}=\Tr(-1)^Fe^{-\beta \Delta}t^{2J_3-R}, \notag
\end{align}
where $\Delta=(Q)^2$.}
To obtain the refined index, we need a ``gauging'' of this global R-symmetry, which modifies the moduli space of the theory by induced mass terms. 
Under the gauged $U(1)_J'$ symmetry with a constant background $A_J=\e$, the $\tau$-derivatives of $Z$ and $\lambda_z$ are modified into
\begin{align}
\del_{\tau}Z&\to (\del_{\tau}+i\e)Z, ~~~~~~~~~~\del_{\tau}\lambda_z\to (\del_{\tau}+i\e)\lambda_z.
\end{align}
This is known to the $\Omega$-background.
Thus, we obtain the modified BRST transformations:
\be
\begin{array}{lcl}
Q_\e Z=i\lambda_{z}, &&Q_\e \lambda_{z}=i(\Dcal_{\tau}Z+[\sigma ,Z]+i\e Z), \\
Q_\e\Zbar=-i\lambda_{\zbar}, && Q_\e \lambda_{\zbar}=-i(\Dcal_{\tau}\Zbar +[\sigma,\Zbar]-i\e \Zbar),\\
Q_\e A=i\eta, &&\\
Q_\e \sigma=\eta, && Q_\e \eta=-\Dcal_{\tau}\sigma,\\
Q_\e Y_{\R}=i(\Dcal_{\tau} \chi_{\R} +[\sigma, \chi_{\R}]), && Q_\e \chi_{\R}=iY_{\R}.
\end{array}
\label{BRST transformations}
\ee
The BRST transformations are nilpotent up to the time translation, gauge transformation including the gauged $U(1)_J'$ transformation.

The action of the modified theory is obtained  
by replacing simply $Q$ with $Q_\e$ in (\ref{Q-exact action}).

\subsection{Chiral multiplet}

Let us now construct the theory which includes a chiral multiplet.
The chiral multiplet is composed of a complex scalar $q$, two complex fermions $\psi_\alpha$, and an auxiliary complex scalar $F$. In this paper, we only consider   chiral multiplets in the fundamental representation. The representations under $SU(2)_J$ and  charges under $U(1)_R$ of the chiral multiplet are summarized in Table \ref{tab:rc}.

\begin{table}[h]
\centering
\begin{tabular}{c||c|c|c} 
~& $q $ & $\psi_{\alpha} $ & $F$ \\ \hline
$SU(2)_J$ & $\mathbf{1}$ & $\mathbf{2}$ &  $\mathbf{1}$ \\ \hline
$U(1)_R$ & $r$ & $r-1$ & $r-2$  \\ 
\end{tabular} 
\caption{The R-symmetries of the chiral multiplet with a  $U(1)_R$ charge $r$.}   
\label{tab:rc}
\end{table}　

The action  is given by
\begin{align}
S_C&=\int dt \,  \Tr \bigg[|\Dcal_0 q|^2 -|X_i q|^2  -i\psibar \sigmabar^0 \Dcal_0 \psi +\psibar \sigmabar^i X_i \psi 
 +|F|^2 +i\sqrt{2}(\qbar \lambda\psi  - \psibar \lambdabar q ) +\qbar D q\bigg].
\end{align}
This action is invariant under the following supersymmetric transformations:
\be
\begin{array}{lcl}
\delta q&=&\sqrt{2}\xi \psi, \\
\delta \psi&=&i\sqrt{2} (\sigma^0 \xibar \Dcal_0 q+i\sigma^i \xibar X_iq)+\sqrt{2}\xi F, \\
\delta F&=& i\sqrt{2}  (\xibar\sigmabar^0\Dcal_0\psi +i \xibar\sigmabar^i X_i \psi)+2i\xibar \lambdabar q.
\end{array}
\ee

After the Wick rotation,  we define the bosonic fields:
\be
\begin{array}{lcl}
Y_\C =F+Zq, && \Ybar_\C=\Fbar+\qbar \Zbar, 
\end{array}
\ee
and the fermionic fields:
\be
\begin{array}{lcl}
\psi =\psi_2, && \psibar=\psibar_2,\\
\chi_\C =-\psi_1, && \chibar_\C=-\psibar_1.
\end{array}
\ee
 The mass dimensions of the fields in the chiral multiplet are
\begin{align}
[q]&=[\qbar]=-\frac{1}{2}, \notag \\
[\psi]&=[\psibar]=[\chi_{\C}]=[\chibar_{\C}]=0,  \\
[Y_{\C}]&=[\Ybar_{\C}]=\frac{1}{2}. \notag 
\end{align}
These fields transform under the BRST symmetry by
\be
\begin{array}{lcl}
Qq =i\psi,&& Q\psi=i(\Dcal_{\tau} q+\sigma q),\\
Q\qbar =-i\psibar, && Q\psibar=-i(\Dcal_{\tau}\qbar-\qbar \sigma),\\
QY_{\C} =i(\Dcal_{\tau}\chi_{\C}+\sigma \chi_{\C}), && Q\chi_{\C,}=iY_{\C},\\
Q\Ybar_{\C} =i(\Dcal_{\tau}\chibar_{\C}-\chibar_{\C}\sigma), && Q\chibar_{\C}=i\Ybar_{\C}.
\end{array}
\ee
Using the BRST charge, the Euclidean action can be written as the $Q$-exact form:
\be
S_C=\frac{1}{2}Q \int d\tau \, \Tr \bigg[\psi\overline{Q\psi}+\psibar \overline{Q\psibar}
-\chi_{\C}\overline{Q\chi_{\C}}-\chibar_{\C} \overline{Q\chibar_{\C}} -2i\chibar_{\C} \mu_{\C}-2i\chi_{\C}\mubar_{\C}\bigg],
\ee
where
\be
\begin{array}{lcl}
\mu_{\C} &=& Zq, \\
\mubar_{\C} &=& \qbar \Zbar,
\end{array}
\label{complex moment maps}
\ee
are (complex) moment map constraints associated with the F-term constraints.
By including the chiral multiplet, the real moment map (D-term constraints)  are also modified to
\be
\mu_{\R} = \frac{1}{2}[Z,\Zbar] + g^2 (q \qbar -\zeta).
\label{real moment maps}
\ee

After these redefinitions of the fields in the chiral multiplet, the theory possesses the following twisted $U(1)_J'$ R-transformation:
\begin{align}
&q\to e^{-i\frac{r}{2}\theta_J}q,~~~~~~~~~~~~~~~~\psi\to e^{-i\frac{r}{2}\theta_J}\psi, \notag \\
&Y_{\C}\to e^{i(1-\frac{r}{2})\theta_J}Y_{\C},~~~~~~~~~\chi_{\C}\to e^{i(1-\frac{r}{2})\theta_J}\chi_{\C}.
\end{align}
We also have the following  $U(1)_R'$ transformation:
\be
\begin{array}{lcl}
q \to e^{ip\theta_R}q, && \psi\to e^{ip\theta_R}\psi, \\
Y_\C \to e^{ip\theta_R}Y_\C, && \chi_\C\to e^{ip\theta_R}\chi_\C,
\end{array}
\ee
where $p$ is a real number. This symmetry is just a flavor symmetry. As similar as the previous section, we gauge these R-symmetries in the constant backgrounds $A_J=\e$ and $A_R=\et$.
The BRST transformations are deformed by
\be
\begin{array}{lcl}
Q_\e q =i\psi,&& Q_\e \psi=i(\Dcal_{\tau} q+\sigma q +i(-\frac{r}{2}\e+p\et) q),\\
Q_\e \qbar =-i\psibar, && Q_\e \psibar=-i(\Dcal_{\tau}\qbar-\qbar \sigma -i(-\frac{r}{2}\e+p\et) \qbar),\\
Q_\e Y_{\C} =i(\Dcal_{\tau}\chi_{\C}+\sigma \chi_{\C}+i((1-\frac{r}{2})\e+p\et)\chi_{\C}), && Q_\e \chi_{\C,}=iY_{\C},\\
Q_\e \Ybar_{\C} =i(\Dcal_{\tau}\chibar_{\C}-\chibar_{\C}\sigma-i((1-\frac{r}{2})\e+p\et)\chibar_{\C}), && Q_\e \chibar_{\C}=i\Ybar_{\C}.
\end{array}
\ee
Here, we have used the same symbol $Q_\e$ as that in (\ref{BRST transformations}), but $Q_\e$ is regarded as including the whole gauged  $U(1)_J'\times U(1)_R'$ symmetries in the following.
Thus, the BRST transformations  are now nilpotent up to the time translation, gauge transformation including gauged $U(1)_J'\times U(1)'_R$ transformations.
Since $p$ is an arbitrary constant, we take 
\begin{align}
p\et=\frac{r}{2}\e +r\et.
\end{align}
Then, the BRST transformation becomes
\be
\begin{array}{lcl}
Q_\e q =i\psi,&& Q_\e \psi=i(\Dcal_{\tau} q+\sigma q +ir\et q),\\
Q_\e \qbar =-i\psibar, && Q_\e \psibar=-i(\Dcal_{\tau}\qbar-\qbar \sigma -ir\et \qbar),\\
Q_\e Y_{\C} =i(\Dcal_{\tau}\chi_{\C}+\sigma \chi_{\C}+i(\e+r\et)\chi_{\C}), && Q_\e \chi_{\C,}=iY_{\C},\\
Q_\e \Ybar_{\C} =i(\Dcal_{\tau}\chibar_{\C}-\chibar_{\C}\sigma-i(\e+r\et)\chibar_{\C}), && Q_\e \chibar_{\C}=i\Ybar_{\C}.
\end{array}
\ee

\section{MPS formula} \label{sec:MPS}
The MPS formula is a formula for computing the total refined index $\Omega(\gamma,y)$ of  multi-centered BPS solutions  with total charge $\gamma$ in terms of the refined indices $\Omega^S(\alpha_i,y)$ for single-centered BPS solutions with  charge $\alpha_i$, where $\gamma=\sum_i \alpha_i$ \cite{Manschot:2010qz,Manschot:2011xc}.   The MPS formula is given by
\begin{align}
\Omegabar(\gamma,y)=\sum_{\substack{\{\alpha_i \in \Gamma \} \\ \sum_i\alpha_i=\gamma  }}\frac{1}{\mathrm{Aut(\{ \alpha_i\})}}g(\alpha_1,\cdots,\alpha_n;y)\Omegabar^S(\alpha_1,y)\cdots \Omegabar^S(\alpha_n,y), \label{eq:MPSformula}
\end{align}
where $\Gamma$ is the charge lattice, $\mathrm{Aut}(\{ \alpha_i\})$ is the symmetry factor appropriate for Maxwell Boltzmann statistics, and
\begin{align}
\Omegabar(\gamma,y)=\sum_{m|\gamma}\frac{y-y^{-1}}{m(y^m-y^{-m})}\Omega(\gamma/m,y^m), \label{eq:rationalinv}
\end{align}
is  the rational invariant.\footnote{The sum runs over all positive integers $m$ such that $\gamma/m$ lies in $\Gamma$.} The function $g(\alpha_1,\cdots,\alpha_n;y)$ is given by
\begin{align}
 g(\alpha_1,\cdots,\alpha_n;y)=(-1)^{\sum_{i<j}\alpha_{ij}+n-1}\bigg[(y-y^{-1})^{1-n}\sum_ps(p)y^{\sum_{i<j}\alpha_{ij}\sign(z_j-z_i)}\bigg], \label{eq:mpsg}
\end{align}
where $\alpha_{ij}=\langle \alpha_i,\alpha_j \rangle$ is the Dirac-Schwinger-Zwanziger (DSZ) product between $\alpha_i$ and $\alpha_j$, and the sum is taken over all collinear solutions to the  equilibrium conditions for the multi-centered BPS  solutions. The $s(p)$ takes $\pm 1$,  which depends on each collinear configuration $p$.

Let us consider the case $\gamma =M\gamma_1 +N\gamma_2$, where $\gamma_{1}$ and $\gamma_2$ are the primitive charge vectors. We denote $\Omega(M\gamma_1+N\gamma_2,y)$ by $\Omega(M,N;y)$ for short. For $(M,N)=(1,1), (1,2), (1,3)$, the MPS formula takes the following expressions  \cite{Manschot:2010qz}:
\begin{align}
\Omegabar(1,1;y)&=[\gamma_{12}]_{-y}\Omegabar^S(1,0;y)\Omegabar^S(0,1;y),  \label{eq:MPS1}  \\
\Omegabar(1,2;y)&=[2\gamma_{12}]_{-y}\Omegabar^S(0,2;y)\Omegabar^S(1,0;y)+\frac{1}{2}[\gamma_{12}]_{-y}^2[\Omegabar^S(0,1;y)]^2\Omegabar^S(1,0;y) \notag \\
&+[\gamma_{12}]_{-y}\Omegabar^S(0,1;y)\Omegabar^S(1,1;y), \label{eq:MPS2} \\
\Omegabar(1,3;y)&=[3\gamma_{12}]_{-y}\Omegabar^S(0,3;y)\Omegabar^S(1,0;y)+[2\gamma_{12}]_{-y}\Omegabar^S(0,2;y)\Omegabar^S(1,1;y) \notag \\
&+[\gamma_{12}]_{-y}\Omegabar^S(0,1;y)\Omegabar^S(1,2;y)+[\gamma_{12}]_{-y}[2\gamma_{12}]_{-y}\Omegabar^S(0,1;y)\Omegabar^S(0,2;y)\Omegabar^S(1,0;y) \notag \\
&+\frac{1}{2}[\gamma_{12}]_{-y}^2[\Omegabar^S(0,1;y)]^2\Omegabar^S(1,1;y)+\frac{1}{6}[\gamma_{12}]_{-y}^3[\Omegabar^S(0,1;y)]^3\Omegabar^S(1,0;y), \label{eq:MPS3} 
\end{align}
where
\begin{align}
[x]_y&=\frac{y^x-y^{-x}}{y-y^{-1}}.
\end{align}

\end{document}